\documentclass[iop]{emulateapj}
\usepackage{apjfonts}
\usepackage{psfig}
\usepackage{mathrsfs}
\usepackage{amsmath}
\usepackage{hyperref}
\usepackage{cleveref}
\usepackage{natbib}
\begin{document}

\slugcomment{AJ, in press}
\shorttitle{NGC\,6402. II. Variable stars}
\shortauthors{C. Contreras Pe\~{n}a et al.}

\title{THE GLOBULAR CLUSTER NGC\,6402 (M14). II. VARIABLE STARS\footnotemark[1]\footnotetext[1]{Based on observations obtained with the 0.9m and 1m telescopes at the Cerro Tololo Interamerican Observatory, Chile, operated by the SMARTS consortium.}}

\author{C. Contreras Pe\~{n}a,\altaffilmark{2,3,4} M. Catelan,\altaffilmark{2,5} F. Grundahl,\altaffilmark{3}
        A. W. Stephens,\altaffilmark{6} \and H. A. Smith\altaffilmark{7}}

\altaffiltext{2}{Pontificia Universidad Cat\'olica de Chile, Instituto de 
       Astrof\'\i sica, Av. Vicu\~{n}a Mackenna 4860, 
       782-0436 Macul, Santiago, Chile; e-mail: mcatelan@astro.puc.cl}

\altaffiltext{3}{Danish AsteroSeismology center (DASC), Department of Physics and Astronomy, Aarhus University, DK-8000 Aarhus C, Denmark}

\altaffiltext{4}{School of Physics, Astrophysics Group, University of Exeter, Stocker Road, Exeter EX4 4QL, UK; e-mail: c.contreras@exeter.ac.uk}

\altaffiltext{5}{Millennium Institute of Astrophysics, Santiago, Chile}

\altaffiltext{6}{Gemini Observatory, 670 North A'ohoku Place, Hilo, HI 96720, USA}

\altaffiltext{7}{Department of Physics and Astronomy, Michigan State University, East Lansing, MI 48824}

\begin{abstract}
We present time-series $BVI$ photometry for the Galactic globular cluster NGC\,6402 (M14). The data consists of $\sim$137 images per filter, obtained using the 0.9m and 1.0m SMARTS telescopes at the Cerro Tololo Inter-American Observatory. The images were obtained during two observing runs in 2006-2007. The image-subtraction package ISIS, along with DAOPHOT\,II/ALLFRAME, were used to perform crowded-field photometry and search for variable stars. We identified 130 variables, 8 of which are new discoveries. The variable star population is comprised of 56 ab-type RR Lyrae stars, 54 c-type RR Lyrae, 6 type II Cepheids, 1 W~UMa star, 1 detached eclipsing binary, and 12 long-period variables. We provide Fourier decomposition parameters for the RR Lyrae, and discuss the physical parameters and photometric metallicity derived therefrom. The M14 distance modulus is also discussed, based on different approaches for the calibration of the absolute magnitudes of RR Lyrae stars. The possible presence of second-overtone RR Lyrae in M14 is critically addressed, with our results arguing against this possibility. By considering all of the RR Lyrae stars as members of the cluster, we derive $\langle P_{ab}\rangle = 0.589$~d. This, together with the position of the RR Lyrae stars of both Bailey types in the period-amplitude diagram, suggests an Oosterhoff-intermediate classification for the cluster. Such an intermediate Oosterhoff type is much more commonly found in nearby extragalactic systems, and we critically discuss several other possible indications that may point to an extragalactic origin for this cluster. 
  
\end{abstract}


\section{Introduction}\label{sec:intro}

NGC\,6402 (M14, $\alpha$ = 17:37:36.1, $\delta$ = -03:14:45, J2000, $\ell = 21.32\degr$, $b = 14.81\degr$) is a moderately high-metallicity globular cluster (GC), with ${\rm [Fe/H]} = -1.28\pm0.08$~dex in the UVES scale of \citet{cg2}, as extensively discussed by \citet{2013Contreras}. Located at a distance of 9.3~kpc from the Sun and 4.0~kpc from the Galactic center, it is also the 10th brightest Galactic GC, with $M_{V} = -9.1$~mag;\footnote{Unless otherwise stated, cluster parameters were taken from the 2010 edition of the \citet{h96,wh10} catalog.} however, its fairly high reddening, $E(\bv)=0.57\pm0.02$~mag \citep{2013Contreras}, probably accounts for the fact that it has been fairly neglected in the literature. 

In \citet{2013Contreras}, we presented the deepest CMD of M14 to date, based on which we investigated several photometric parameters of the cluster, such as metallicity, foreground reddening, age, and horizontal branch (HB) morphology. That study confirms the unusual HB morphology of M14, with stars extending through the RR Lyrae gap to an extended blue tail reaching the main-sequence turnoff level. The HB morphology is considerably bluer than  expected for the cluster's metallicity, which makes M14 a second-parameter GC \citep[see][for a review]{2009Catelan}.


M14 has been long studied in search of variable stars, and it is indeed known to contain a large number of variables, including RR Lyrae stars and type II Cepheids (references and a summary of the variable stars in M14 can be found in \citealt{1994We_Fro}, hereafter WF94, and \citealt{2012Conroy}, hereafter CDLM12). It is also one of the few clusters where a nova eruption has been observed \citep{nova1938,sharam14}. 

A recent review of the properties of RR Lyrae stars is provided in the monograph by \citet{CS15}. They correspond to HB stars crossing the classical instability strip. These variables present radial pulsations with periods in the range $0.2 \lesssim P \lesssim 1$~d, with the mode of pulsation typically being either the fundamental (\textit{RRab} or RR0, $P \gtrsim 0.4$~d) or the first overtone (\textit{RRc} or RR1, $P \lesssim 0.4$~d). The first group can present amplitudes of variation up to 1.3~mag in $V$, with amplitude decreasing (on average) with increasing period, while the second shows fairly uniform amplitudes of $\sim 0.4$~mag in $V$, but typically decreasing towards both the long and short period ends. Differences between the two groups are easily observed when comparing their light curves, with RRc stars typically presenting much more sinusoidal variations, and also show up in the period-amplitude plane (also known as \textit{Bailey diagram}). In addition, a group of RR Lyrae stars, known as \textit{RRd} (or RR01) \textit{stars}, has been found to pulsate simultaneously in both the fundamental mode and the first overtone, with typical period ratios in the range $0.73 < \langle P_{1}/P_{0} \rangle < 0.75$ \citep[e.g.,][and references therein]{1983Cox,1985Nemec,2000Alcock,2009Catelan}. More recently, additional sequences have been identified in the period ratio vs. period (Petersen) diagram, likely due to the presence of non-radial modes
\citep[e.g.,][]{2016Netzel,2017Smolec}. 

It has been suggested that the second overtone may also be excited in some RR Lyrae stars, so-called RRe (or RR2) stars. These would be short-period, low-amplitude variables presenting asymmetric light curves \citep{1996Alcock}, representing a different group in Bailey diagrams \citep[see, e.g.,][]{2000Clement} and located close to the blue edge of the instability strip, which would quench the amplitude of oscillation \citep{1996Walker}. \citet{1996Alcock}, in a study of the properties of RR Lyrae stars in the LMC, found three peaks in the period distribution of such stars, at periods 0.59, 0.34, and 0.28~d which they tentatively associated to RRab, RRc, and RRe variables respectively. However, several authors \citep[e.g.,][]{kovacs1998,2004Catelanb} have argued that the third group may simply represent the short-period tail of the RRc variables distribution, rather than a different pulsation mode.

When studying the properties of  RR Lyrae stars in a sample of 5 GCs ($\omega$~Cen~= NGC\,5139, M3~= NGC\,5272, M5~= NGC\,5904, M15~= NGC\,7078, M53~= NGC\,5024), \citet{1939Oosterhoff} realized that there was a clear division between the clusters, in terms of the mean periods of ab- and c-type variables, dividing them in two groups. The so-called Oosterhoff type I clusters (OoI), like M3 and M5, showed a mean period of their RRab stars $\langle P_{ab}\rangle \approx 0.55$~d, with a low number of RRc variables. Oosterhoff type II clusters (OoII), such as $\omega$~Cen, M15, and M53, in turn, showed $\langle P_{ab} \rangle \approx 0.65$~d, with a higher fraction of RRc stars compared to ab-type pulsators. Subsequent works confirmed this tendency for GCs in the Milky Way. Confirming the original findings by \citet{1955Arp} and \citet{1959Kinman}, it is now clear that both groups also differ in metallicity, with OoI clusters being, on average, more metal-rich than OoII clusters \citep[see][for recent reviews and references]{2009Catelan,CS15}. 

In the Milky Way, GCs show a clear division between OoI and OoII groups in the $\langle P_{ab} \rangle$-metallicity plane, with very few Galactic globulars falling in the period range $0.58 < \langle P_{ab} \rangle < 0.62$~d,  also known as the ``Oosterhoff gap'' \citep{2004Catelanb}.  In addition, two GCs, NGC\,6441 and NGC\,6388, appear to define a third group (OoIII), based on their unusually high $\langle P_{ab} \rangle$ values for their metallicities \citep{2000Pritzl}. Additional subclassifications have also on occasion been suggested \citep[e.g.,][]{CQ87,YL02}.

The Oosterhoff dichotomy seems to be present only for Galactic GCs, as dwarf spheroidal (dSph) satellite galaxies and their respective GCs fall preferentially on the Oosterhoff gap region, thus being called ``Oosterhoff-intermediate'' (OoInt). The same effect is observed in the [Fe/H] vs. HB morphology plane, where extragalactic GCs tend to clump in a narrow area which is avoided by Galactic GCs. This poses some useful constraints regarding the formation history of the Galactic halo, since it strongly suggests that the latter cannot have been assembled by ``protogalactic fragments'' similar to the early counterparts of most of the present-day Milky Way dSph satellites, or else the Oosterhoff dichotomy would not exist \citep{2009Catelan}. However, this does not exclude the possibility that there may have existed primordial dwarf galaxy(ies) with properties unlike those of the dSph galaxies we observe today, and indeed the accretion of extragalactic GCs has been frequently suggested in the literature as playing an important role in the build-up of the present-day Milky Way halo \citep[e.g.,][]{1978Sea_Zinn, YL02, 2007Lee, 2007Gao, 2010Forbes, 2010Carretta}. In this sense, at least some of the so-called ``ultra-faint'' dSph galaxies do harbor RR Lyrae stars whose properties are consistent with those of the most metal-poor components of the Galactic halo \citep[][and references therein]{2014Clementini, 2016Vivas}. For recent discussions on the possible astrophysical origins of the Oosterhoff dichotomy, we refer the reader to \citet{asea14}, \citet{jl15}, and \citet{dvea16}. 

WF94 originally classified M14 as an OoI cluster, as expected for its metallicity. On the other hand, GCs with blue HB morphologies are normally found to be OoII clusters, and as we have already mentioned, M14 does contain an extended blue HB tail \citep{2013Contreras}. In fact, M14, along with M62 (NGC\,6266) and NGC\,4147, have been suggested to belong to an exclusive group of OoI clusters having predominantly blue HBs \citep{rcon1}. Thus, studying the properties of the RR Lyrae population of M14 has the potential of disclosing the physical reason for the existence of the Oosterhoff dichotomy.

It is the purpose of this work to search for variable stars utilizing modern image-subtraction techniques, which have proven to be very successful in the crowded areas of GCs, thus potentially allowing us to detect a large number of previously unknown variable stars down to the core of M14. Our paper is structured as follows. In Sect.~\ref{sec:obs} we describe our dataset and the procedures adopted in our analysis. In Sect.~\ref{sec:vars} we report on our detection of variable stars in M14, and describe their general properties. In Sect.~\ref{sec_fourier} we provide the results of our Fourier decomposition of M14's RR Lyrae stars, and derive the physical parameters of these stars based on calibrations available in the literature. The distance modulus of the cluster is derived in Sect.~\ref{sec:DM}, while the status of the candidate second-overtone RR Lyrae stars is critically discussed in Sect.~\ref{sec:rre}. The Oosterhoff type of M14 is obtained in Sect.~\ref{sec:oost}, where a possible extragalactic origin of the cluster is also addressed. Finally, our results are summarized in Sect.~\ref{sec:summary}.

\section{Observations and Data Reduction} \label{sec:obs}

Time-series $B$, $V$, and $I$ photometry was obtained for M14 using two different telescopes. The first set was obtained with the SMARTS 1.0m telescope at Cerro Tololo Interamerican Observatory (CTIO) for 7 consecutive nights between August 10-16 and 5 nights between September 28-October 3 2006, comprising a total of 75, 72, and 80 images in the $B$, $V$, and $I$ filters, respectively. Typical exposure times corresponded to 300~s ($B$), 150~s ($V$), and 100~s ($I$). The typical seeing level was 1.6\arcsec. The telescope was equipped with the Y4KCam $4064\times4064$ CCD detector, with a pixel scale of 0.289\arcsec/pixel, producing a $20\arcmin \times 20\arcmin$ field of view (FOV). The detector is read out in 4 quadrants, each with a different readout noise and gain, varying between 4.78 and 4.98~$e^{-}$ and 1.33 and 1.42~$e^{-}$/ADU, respectively. The raw images look like the field has been cut in 4 pieces by a cross-shaped dead-region, which corresponds to the overscan sections. In order to avoid the boundaries between the quadrants, the cluster was centered on the top left quadrant. Here we only analyze the quadrant containing the cluster. The second set of time-series $B$, $V$, and $I$ photometry was obtained with the SMARTS 0.9m telescope at CTIO during 16 nights between June 22-July 12 2007, comprising a total of 67, 65, and 62 images in $B$, $V$, and $I$, respectively. Typical exposure times of 600~s in $B$, $V$ and 150~s in $I$ were used, and the typical seeing was 1.8\arcsec. A $2048\times2046$ CCD detector with a pixel scale of 0.369\arcsec/pixel, equivalent to a field size of $13.5\arcmin\times 13.5\arcmin$, was used for image acquisition. The detector has a readout noise and gain of 2.7~$e^{-}$ and 0.6~$e^{-}$/ADU, respectively.

\subsection{Search for variable stars}\label{ref:search}

The  image  subtraction  technique  is  one of  the  best  tools  for  identifying  variable  stars  in  crowded fields  like  GCs.  We  have  made use of the  ISIS v2.2  package \citep{2000Alard}  for  this purpose. The $BVI$ photometry of both sets of images were analyzed independently using this software.

Several steps are part of the ISIS reduction procedure. Firstly we get rid of any shifts and rotations between the images by interpolating to a common reference frame using standard IRAF routines.\footnote{IRAF is distributed by the National Optical Astronomy Observatory, which is operated by the Association of Universities for Research in Astronomy (AURA) under a cooperative agreement with the National Science Foundation.} We then create a reference image for the difference imaging process. In order to increase the signal-to-noise ratio as well as to account for cosmic rays or sky variations (although other characteristics were also checked, such as sky background and the saturation level of the brightest stars), we selected the 10 best-seeing images in each filter. The next step, the image subtraction process proper, consists in finding a convolution kernel that will transform the reference image to match any given image, thus accounting for the seeing variations between them, and then subtracting the convolved image from the original one. In the subtracted images, stars of constant brightness should cancel out, and only variable stars leave a residual flux. Then, ISIS constructs a median image of all of the subtracted images, enhancing the variability signal and thus simplifying the variable star detection. Finally, ISIS performs point-spread function (PSF) photometry for variable star candidates on all of the subtracted images. The latter yields an output file which contains the Julian date of observation, the corresponding relative flux and its associated error for every epoch. 

\begin{figure*}
\begin{center}
\resizebox{1.4\columnwidth}{!}{\includegraphics{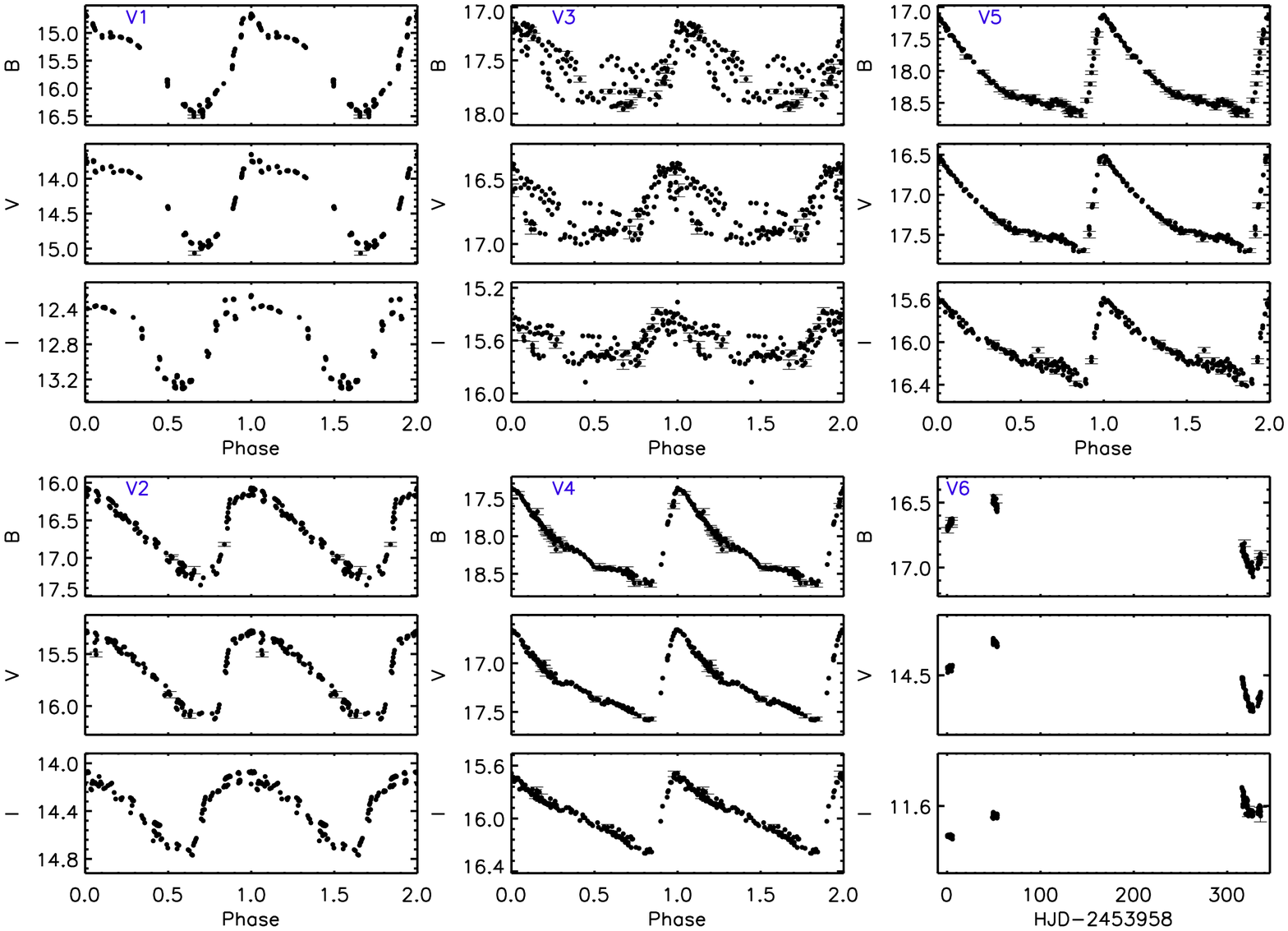}}\\
\resizebox{1.4\columnwidth}{!}{\includegraphics{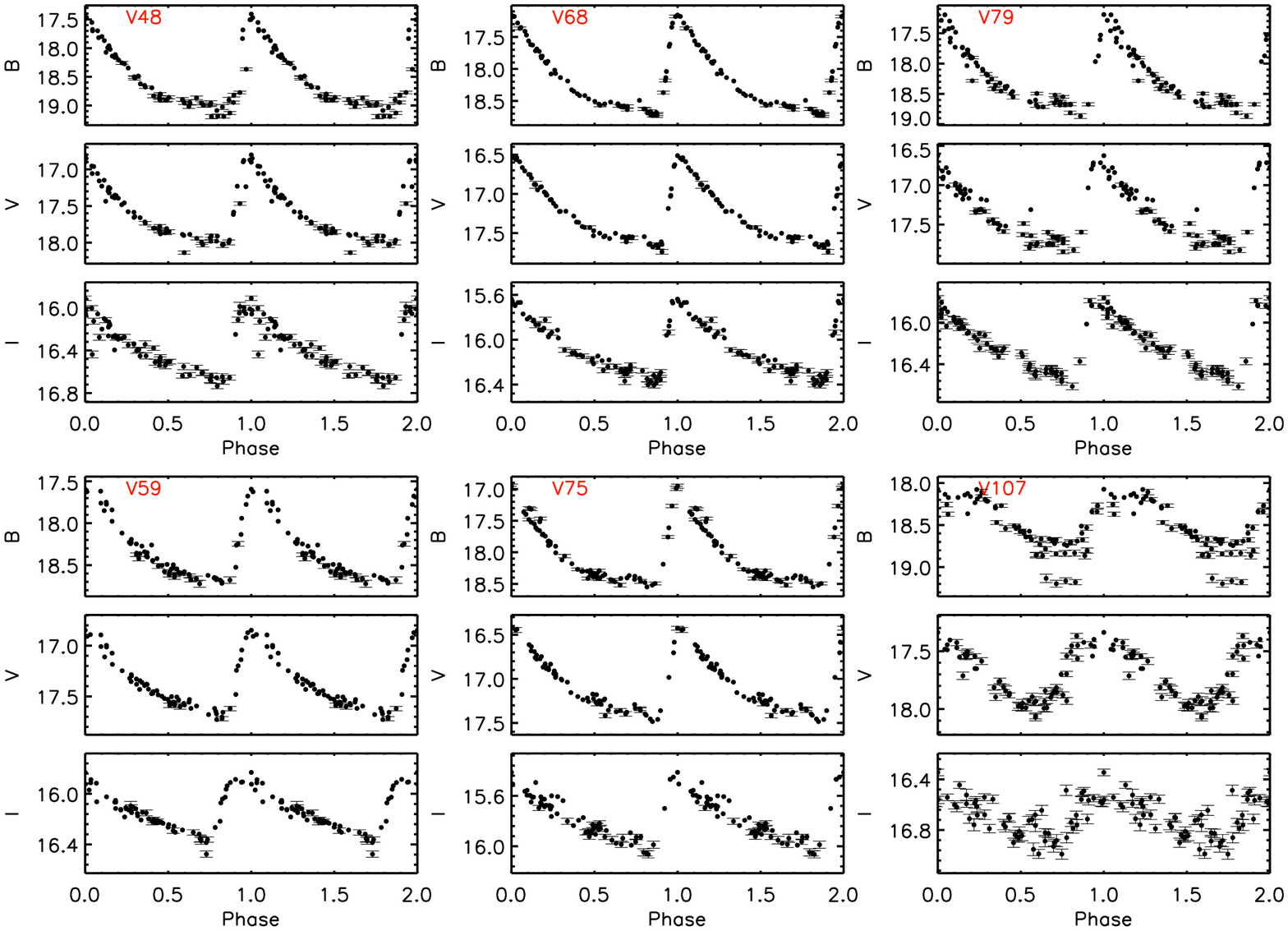}}
\caption{$BVI$ light curves for a sample of variable stars in M14. For clarity, photometric error bars are overplotted for errors larger than $0.02$~mag only. The top six light curves ({\em blue labels}) correspond to objects found outside the core radius of the cluster, while the bottom six light curves ({\em red labels}) depict objects that are located within the central part of M14. Note the trend towards increased error bars and photometric scatter as one moves inwards across the field. The entire set of light curves is available online. }
\label{lcurves}
\end{center}
\end{figure*}

\begin{figure}
\begin{center}
\resizebox{\columnwidth}{!}{\includegraphics{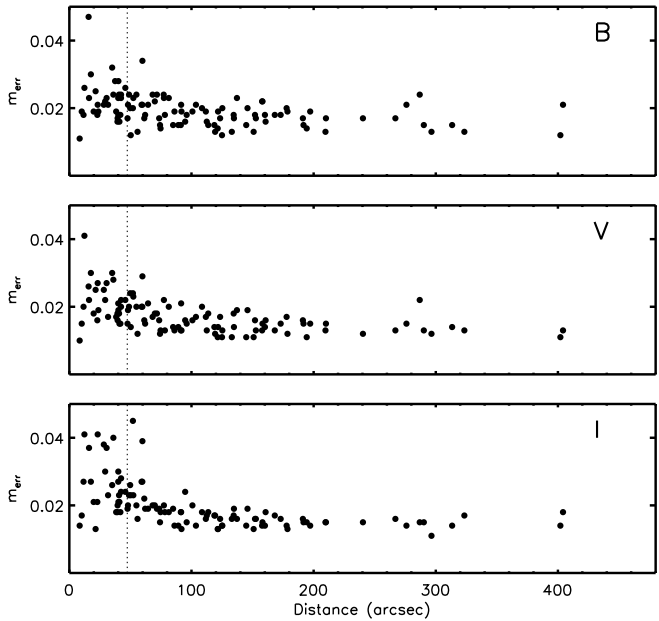}}
\caption{Mean photometric error vs distance from the center of the GC for $B$ (top), $V$ (middle) and $I$ (bottom). The vertical dashed line marks the core radius.}
\label{phot_merr}
\end{center}
\end{figure}

To transform the relative fluxes into calibrated magnitudes, we first followed the procedure described in \citet{2002Moj}. However, we note that the calibration of ISIS relative fluxes has been shown to be uncertain \citep[see, e.g.,][]{2005Bal}. For this reason, we also performed crowded-field PSF photometry on both sets of images, using DAOPHOT\,II/ALLFRAME \citep{1994Stetson}. Standard fields were only observed for the 0.9m run \citep[the calibration of the 0.9m photometry is described in more detail in][]{2013Contreras}. In order to bring the SMARTS 1.0m instrumental magnitudes to the standard Johnson-Cousins system, we performed a cross-correlation between the 0.9m and 1.0m catalogues, selecting stars with small photometric errors, that were located outside the core radius and that covered a wide range in color. We calibrated the 1.0m data by means of least-squares fits, following \citet{1998Montegriffo}. Unfortunately, this procedure was unsuccessful in calibrating the 1.0m data for some of our variable stars, for which a large offset between the two runs is observed. This effect was more evident in stars that were located inside the core radius. When this occurs, we only use data arising from the 0.9m run.  We also note that the accuracy of the light curve decreases as we get closer to the center of the cluster. Figure ~\ref{lcurves} shows a subsample of light curves of M14 variable stars, in which we can observe that stars inside the core radius tend to have noisier light curves and larger photometric errors. Figure~\ref{phot_merr} shows that the mean photometric error increases as we get closer to the centre of the cluster, an effect that is much stronger in $I$. The Heliocentric Julian Day, magnitude and magnitude errors (as provided by DAOPHOT\,II/ALLFRAME) for the variable stars studied in this paper are presented in Table~\ref{lcurves_ascii}.

Finally, we determined the periods of variable stars using the phase dispersion minimization (PDM) algorithm of \citet{pdm1}, implemented in the NOAO ASTUTIL IRAF package.

\begin{deluxetable}{lcccc}
\tabletypesize{\scriptsize} 
\tablecaption{Light curves of variable stars of M14.\tablenotemark{a}}
\tablehead{\colhead{ID} & \colhead{Filter} & \colhead{HJD} & \colhead{Mag} & \colhead{Mag\_err}  
}
\startdata
{V1} & {B} &  {2453958.58210} &  {15.648} &  {0.010} \\
{V1} & {B} &  {2453958.58556} &  {15.604} &  {0.003} \\
{V1} & {B} &  {2453958.59694} &  {15.594} &  {0.003} \\
{V1} & {B} &  {2453958.60832} &  {15.606} &  {0.004} \\
{V1} & {B} &  {2453958.62411} &  {15.585} &  {0.005} \\
{V1} & {B} &  {2453959.57059} &  {15.058} &  {0.005} \\
{V1} & {B} &  {2453959.58197} &  {15.079} &  {0.005} \\
{V1} & {B} &  {2453959.59335} &  {15.044} &  {0.005} \\
{V1} & {B} &  {2453959.60484} &  {15.060} &  {0.004} \\
{V1} & {B} &  {2453959.61622} &  {15.036} &  {0.004} \\
{V1} & {B} &  {2453959.62759} &  {15.047} &  {0.004} \\
{V1} & {B} &  {2453960.64488} &  {14.711} &  {0.004} \\
{V1} & {B} &  {2453960.65627} &  {14.712} &  {0.004} \\
{V1} & {B} &  {2453960.66914} &  {14.703} &  {0.005} \\
{V1} & {B} &  {2453960.68098} &  {14.737} &  {0.004} \\
{V1} & {B} &  {2453961.58309} &  {14.846} &  {0.003} \\
{V1} & {B} &  {2453961.59446} &  {14.852} &  {0.003} \\
{V1} & {B} &  {2453961.60585} &  {14.832} &  {0.003} \\
{V1} & {B} &  {2453961.61723} &  {14.840} &  {0.002} \\
{V1} & {B} &  {2453961.62862} &  {14.866} &  {0.002} \\
{V1} & {B} &  {2453961.64000} &  {14.861} &  {0.002} 
\enddata
\tablenotetext{a}{See the text for a description of the columns. In here we only show a portion of the table for reference regarding its form and content. The entire table is available online.}
\label{lcurves_ascii}
\end{deluxetable}

\section{Variable Stars}\label{sec:vars}

The current (03/2017) version of the \citet{2001Clement} catalog of variable stars in GCs shows 164 variable stars in and around M14.\footnote{\scriptsize\tt http://www.astro.utoronto.ca/$\sim$cclement/cat/C1735m032} These numbers are based on two major studies of variable stars in the cluster. Variable stars from V1 through V93 arise from WF94, whilst objects V94-V164 come from the variability study of CDLM12. In addition, the catalog includes Nova Ophiuchi 1938, a classical nova \citep{sharam14,nova1938}. Here we follow the designation from \citet{2001Clement}.  

\begin{figure*}
\centering
\resizebox{\columnwidth}{!}{\includegraphics{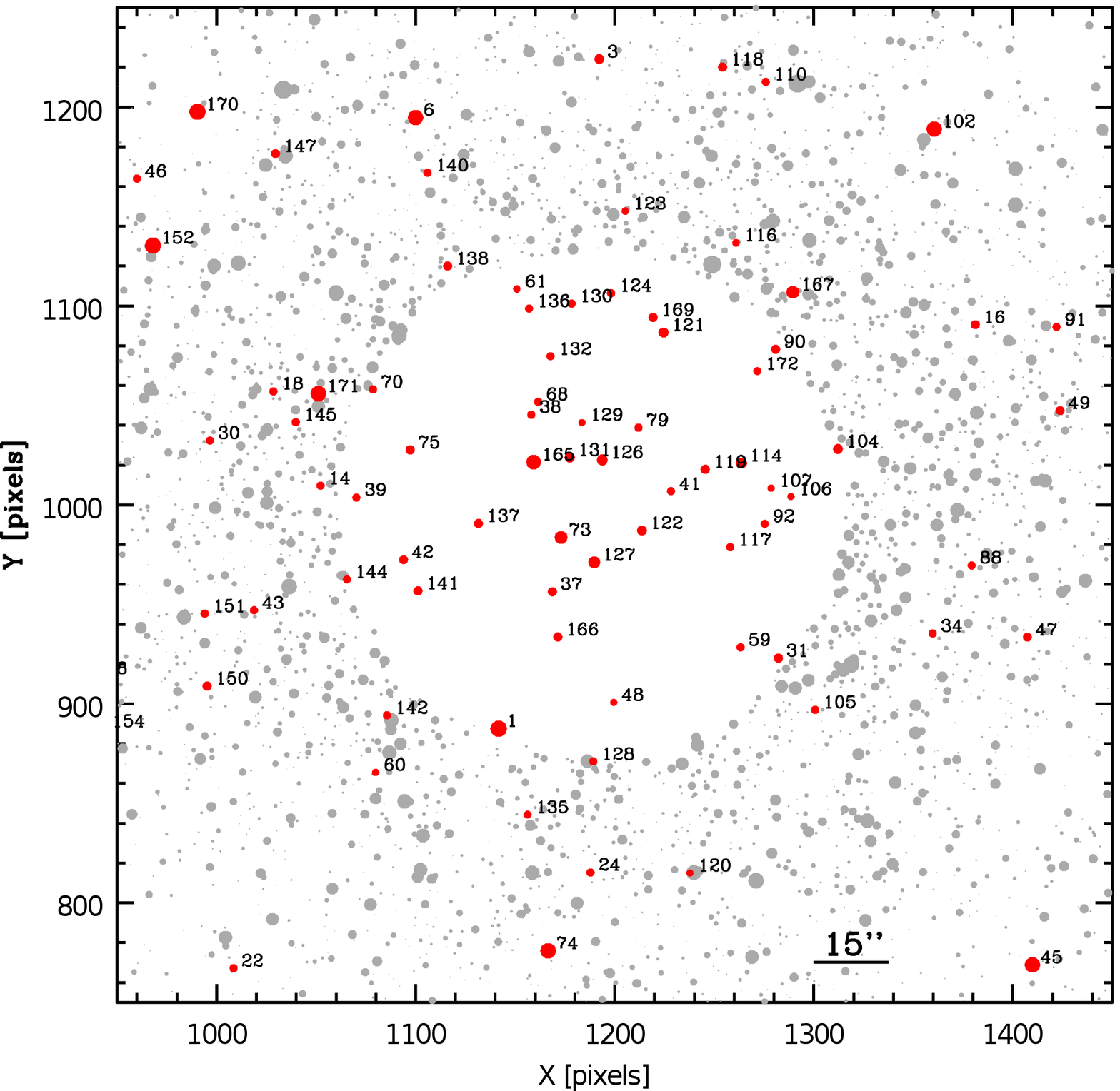}}
\resizebox{\columnwidth}{!}{\includegraphics{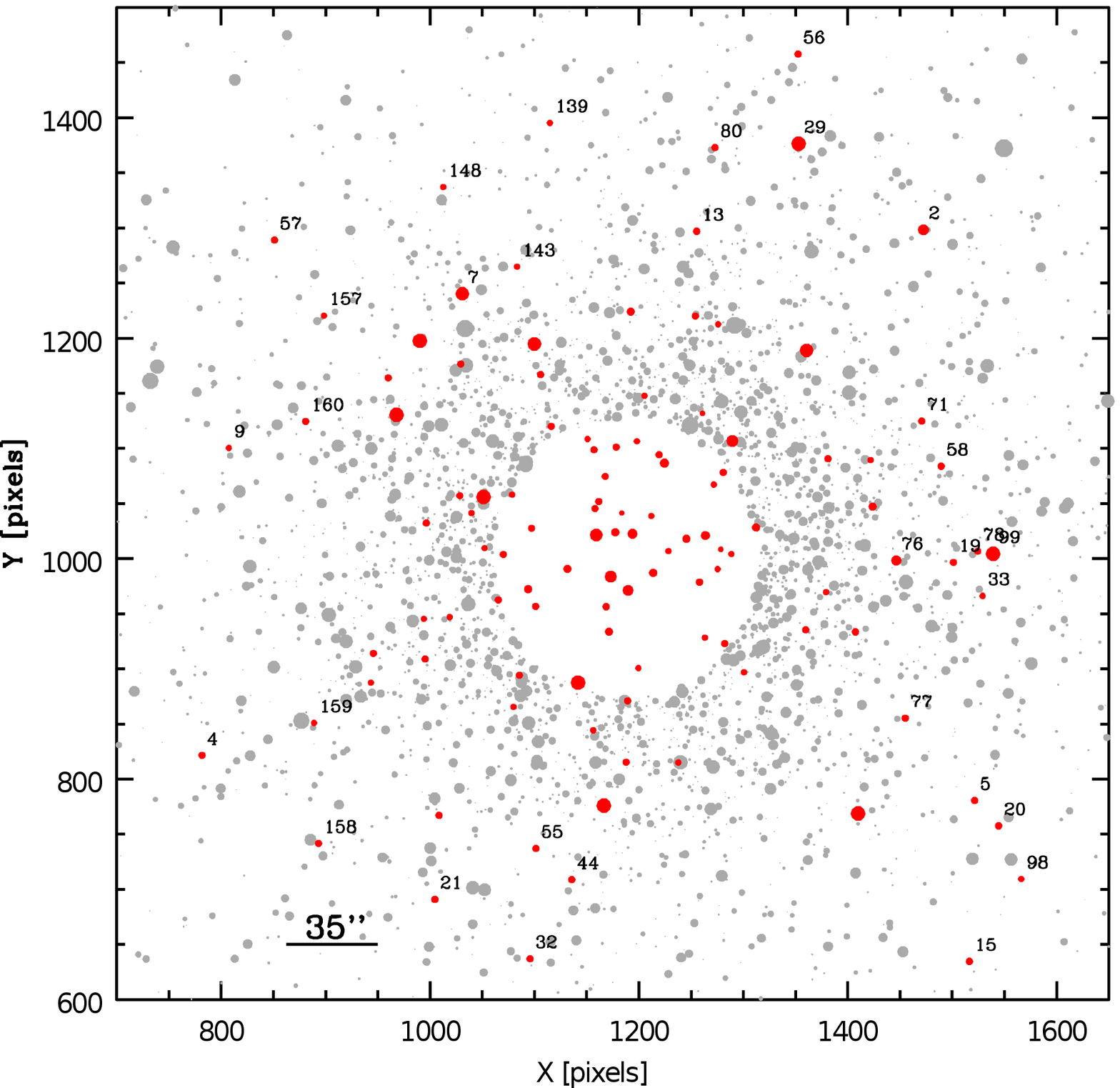}}\\
\resizebox{\columnwidth}{!}{\includegraphics{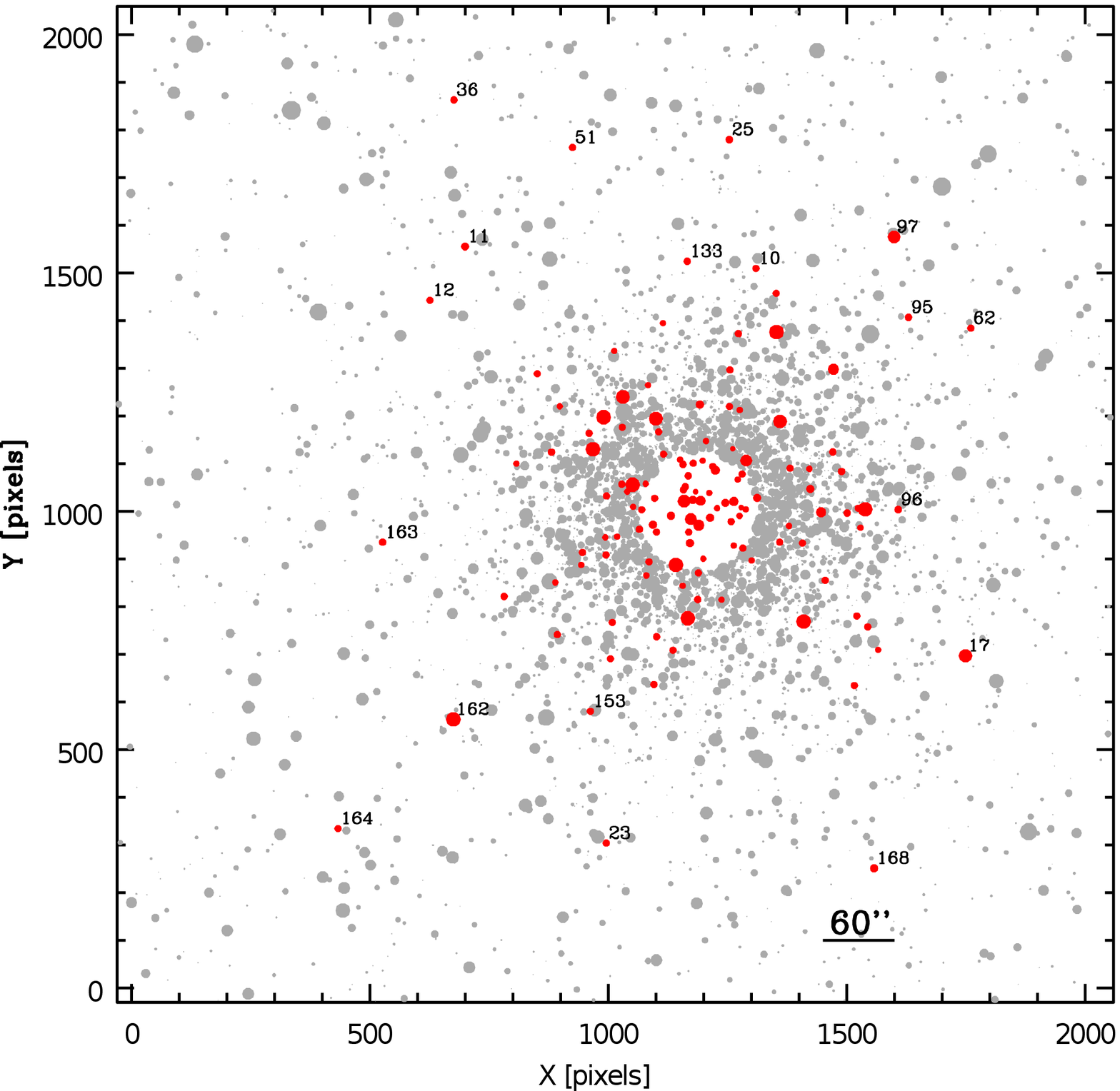}}
\caption{Finding chart showing all of the variable stars of M14. In all of the plots north is down and east is to the left. Constant stars are omitted from the central parts of the cluster, for clarity.}
\label{fchart2}
\end{figure*}

Out of the 93 variable stars from the work of WF94, we are able to recover 68 stars, but do not detect variability for another 22 stars previously reported by these authors. From the remaining three stars, 2 are located outside of our field of view, whilst V50 could correspond to one of the new discoveries (V169, see notes in Table \ref{varnotes}). Out of the 71 stars reported by CDLM12, we detect 54 stars. Among the remaining 17 objects, 12 are not seen to vary in our study, whilst 5 actually correspond to variables from WF94. The Appendix at the end of the paper provides some notes on individual stars, with Table~\ref{varnotes} therein summarizing some of the information on the variables reported in previous studies. We note, in this sense, that we find a different classification compared to previous studies for a single star only. WF94 classifies V77 as an RRc pulsator with a period of 0.327~d; however, our light curve shows that it is instead an ab-type RR Lyrae with a period of 0.7932~d. 

We report the discovery of 8 new variables (V165-V172) in M14, with 3 of them corresponding to RR Lyrae stars, 3 long-period variables (LPVs, or variables with periods longer than our coverage), 1 type II Cepheid and 1 suspected detached eclipsing binary (DEB). The finding chart for the variable stars in M14 is shown in Figure~\ref{fchart2}.

Tables~\ref{varparsa} and \ref{varparsb} show the properties of both new and old variables for this cluster. In Table \ref{varparsa}, column 1 indicates the star's identification, while columns 2 and 3 give the respective right ascension and declination (J2000 epoch). In column 4 we provide our derived periods. Column 5 shows the differential reddening derived in \citet{2013Contreras}, while column 6 gives the classification of each of the variable stars in the cluster. In Table \ref{varparsb}, column 1 indicates the star's identification, while columns 2, 3, and 4 list the amplitude of the light curves in $B$, $V$, and $I$, respectively. Columns 5, 6, and 7 show the derived magnitude-weighted mean $B$, $V$, and $I$ magnitudes, respectively.  Intensity-weighted mean magnitudes are provided in columns 8, 9, and 10. We provide the magnitude- and intensity-weighted mean $B-V$ colors in columns 11 and 12. In the case of RR Lyrae stars, column 13 shows the $B-V$ color corresponding to the equivalent static star (see below).\footnote{For further details on the difference between magnitude- and intensity-weighted means, and how they compare with the corresponding quantities for the so-called ``equivalent static star,'' the reader is referred to \citet{bono1}.} Finally, columns 14 and 15 show the magnitude- and intensity-weighted mean $V-I$ colors.

\begin{figure}
\begin{center}
\resizebox{\columnwidth}{!}{\includegraphics{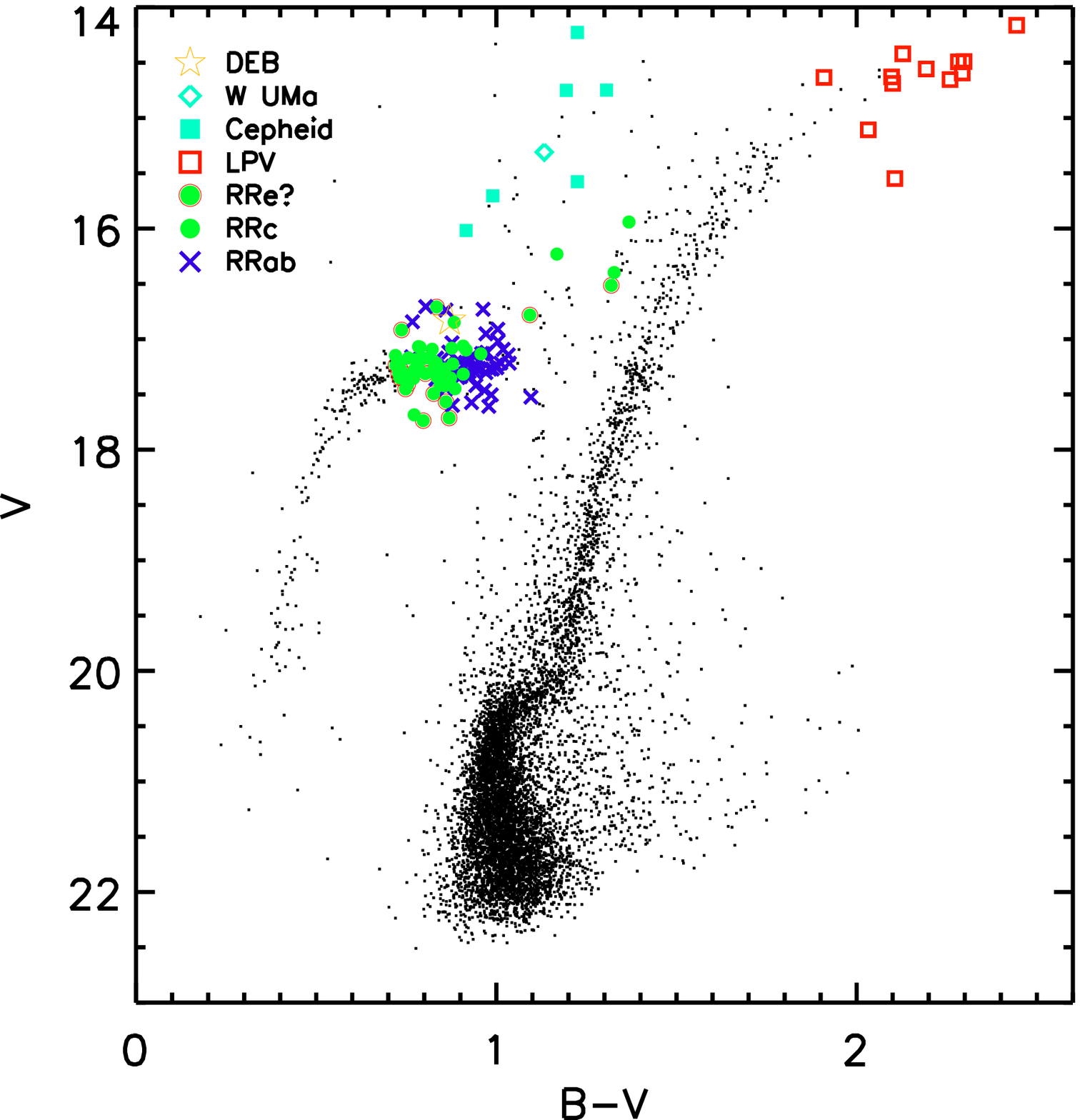}}\\
\resizebox{\columnwidth}{!}{\includegraphics{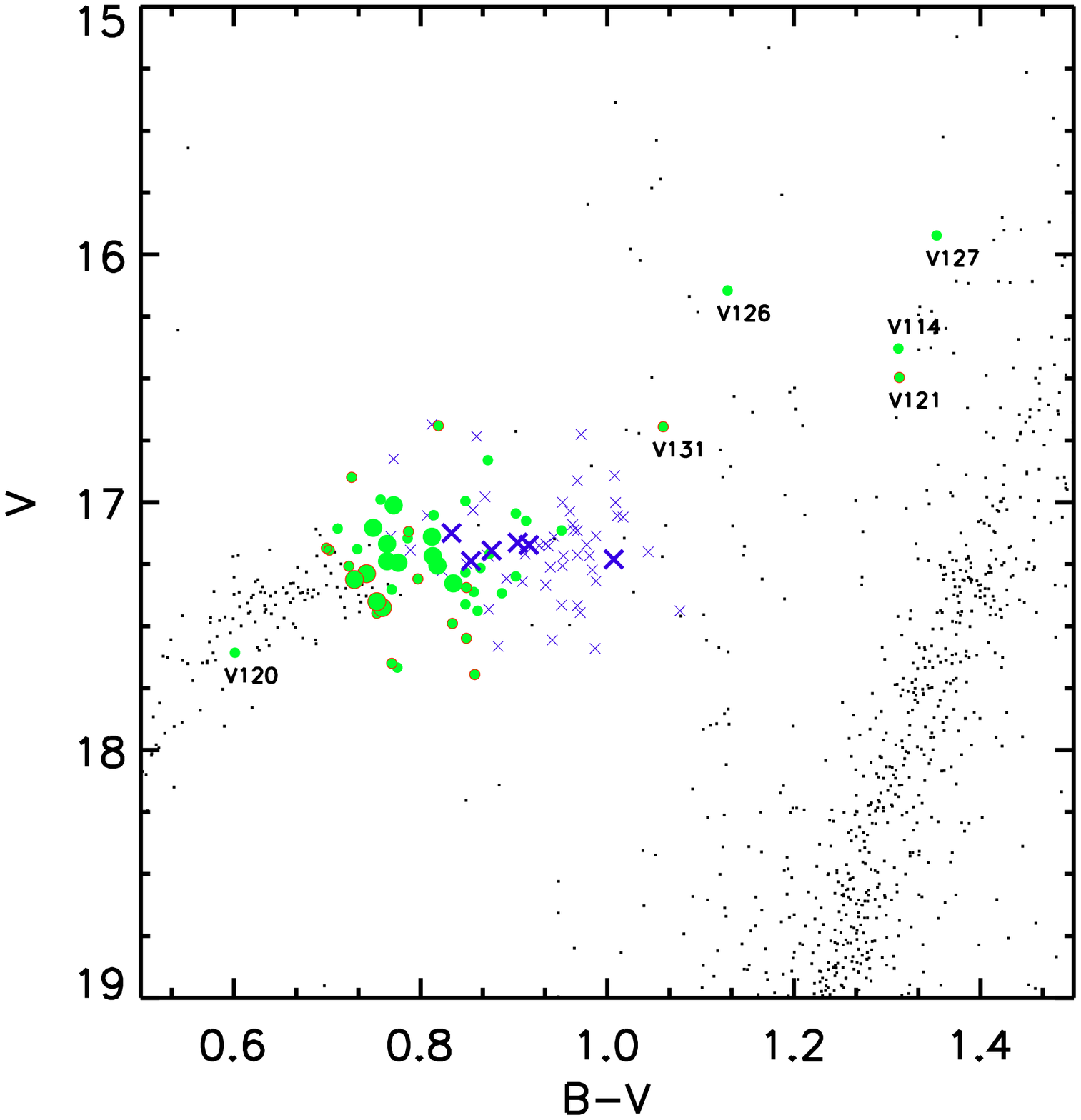}}
\caption{({\em top}) Color-magnitude diagram of M14 \citep[from][]{2013Contreras}, showing the location of the variable stars in M14. RRab and RRc stars are shown as blue crosses and green circles, respectively, whilst candidate RRe stars (see Sect.~\ref{sec:rre}) are also displayed as green circles, but highlighted with thin red outer boundaries. Cyan filled squares denote Type II Cepheids, whereas open red squares indicate the LPVs. A W~UMa system is shown as an open diamond symbol, whereas the DEB is displayed as a golden star. Dots represent non-variable stars. ({\em bottom}) Same as above, but zooming in around the RR Lyrae region. Large symbols mark the RR Lyrae that were used to estimate the distance modulus of M14.}
\label{cmd1}
\end{center}
\end{figure}

Astrometry for each of the variable stars in the cluster was determined by means of an IDL routine, and are tied to the 2MASS system \citep{msea06}. We note that our values for right ascension and declination differ from those presented by CDLM12. We find that our values of right ascension are 3\arcsec~larger, whereas our declination values agree to within 1\arcsec. A similar offset is noted in the 03/2017 version of the \citet{2001Clement} catalogue between the values presented in CDLM12 and those presented by \citet{2009Samus}, with the latter study agreeing with our results. 

In order to determine the color corresponding to the equivalent static star, i.e., the color the star would have if it were not pulsating, we first determine the magnitude-weighted mean color and then applied a correction factor, which is a function of the RR Lyrae blue amplitude (A$_{B}$), taken from Table~4 of \citet{bono1}. This is only done in the case of $\bv$, as the latter authors do not provide correction factors for $V-I$. 

From this point onwards, unless otherwise noted, we will omit the $I$-band photometry from our discussion, since it is considerably noisier than in the bluer bandpasses, and its inclusion does not change any of the main results that we obtain based solely on the $B$ and $V$ data.
 
The location of each of the variables in our final CMD is shown in Fig.~\ref{cmd1}. The positions of the variables are based on the intensity-mean magnitudes, $\langle V \rangle$, and the magnitude-weighted mean $(B-V)$ colors. The RR Lyrae show a rather uniform distribution across the ``gap'' in the HB, and~-- apart from the outliers discussed below~-- a range in magnitudes that is largely consistent with the ``vertical spread'' typically seen at the HB level in the CMDs of other GCs \cite[e.g.,][]{1990Sandage,1992Catelan}. Some RR Lyrae do appear to be fainter than the HB level; however, these are located within the core radius and the apparent faintness is more likely due to uncertain photometry rather than a true effect. There are also other marked exceptions to this behavior, objects V114, V120, V121, V126, V127, and V131. V120 appears to the blue of the instability strip, while the other five stars appear too bright and to the red of the RR Lyrae gap. The photometric uncertainties for these objects do not exceed 0.01~mag. This behavior could be explained by the presence of an unresolved red (or blue, in the case of V120) star companion. The 5 ``red'' stars are located within the core radius, whilst V120 is only 1.26\arcmin~ from the cluster center; thus, blending would not be unexpected in these sources. If stars to the red of the instability strip do in fact have an unresolved red companion, then we would expect to observe a high amplitude ratio, $A_{B}/A_{V}$ \citep[e.g.][]{2001Pritzl}. This is true for V114 and V121, which show  $A_{B}/A_{V} \geq 2$. V127 presents large scatter in its $V$-band light curve, making it difficult to locate the star in our CMD, whilst V131 and V126 do not present a high amplitude ratio or large scatter in $V$. In order to further inspect the possibility of unresolved companions, we checked {\em Hubble Space Telescope} archive images of M14. Objects V114 and V121 are not covered in the images, whilst V120, V126, and V131 clearly showed unresolved companions within 1\arcsec. This is less clear in the case of V127, but at this point we cannot rule out the presence of a faint companion. 

\begin{figure}
\begin{center}
\resizebox{\columnwidth}{!}{\includegraphics{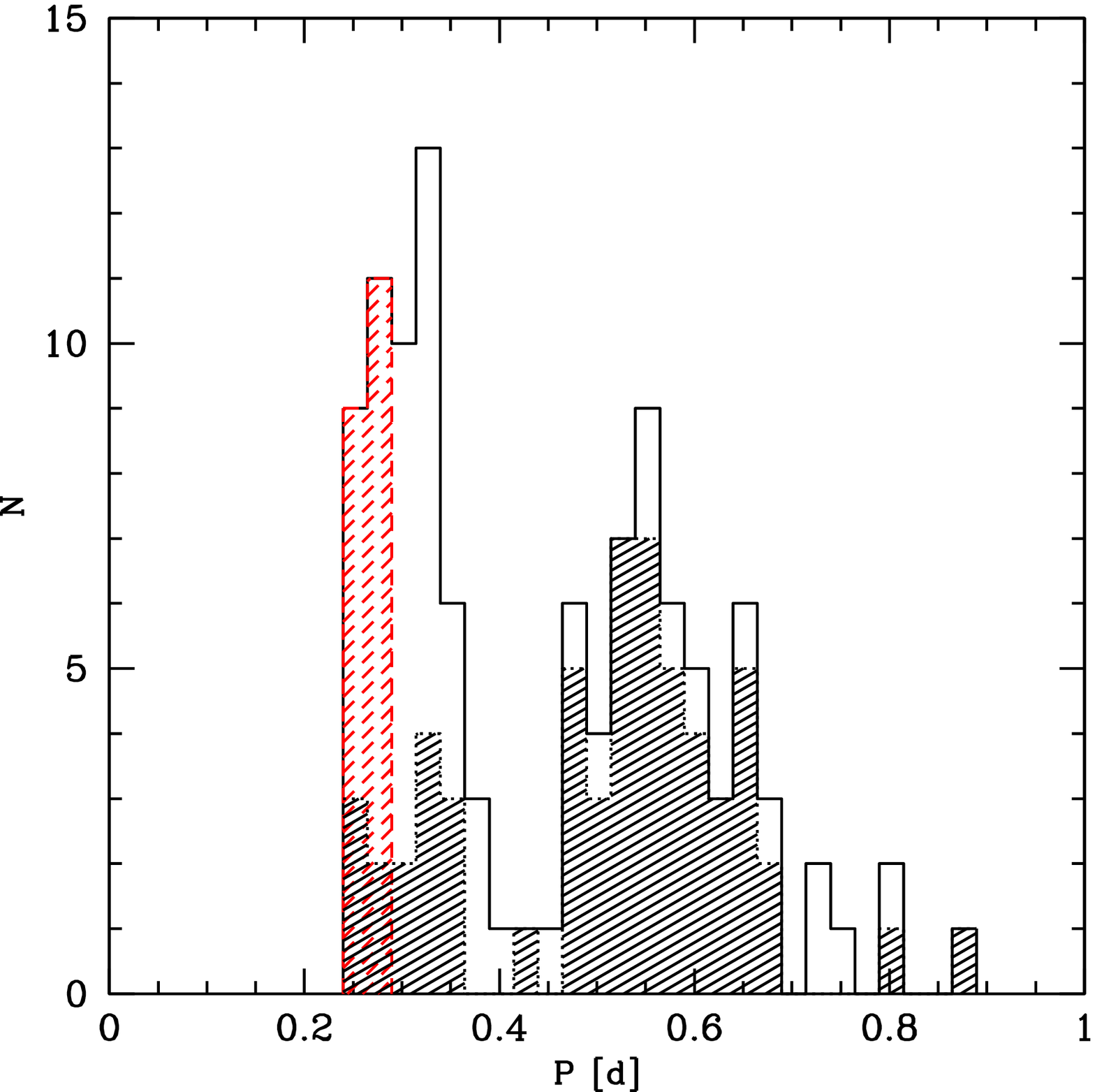}}\\
\resizebox{\columnwidth}{!}{\includegraphics{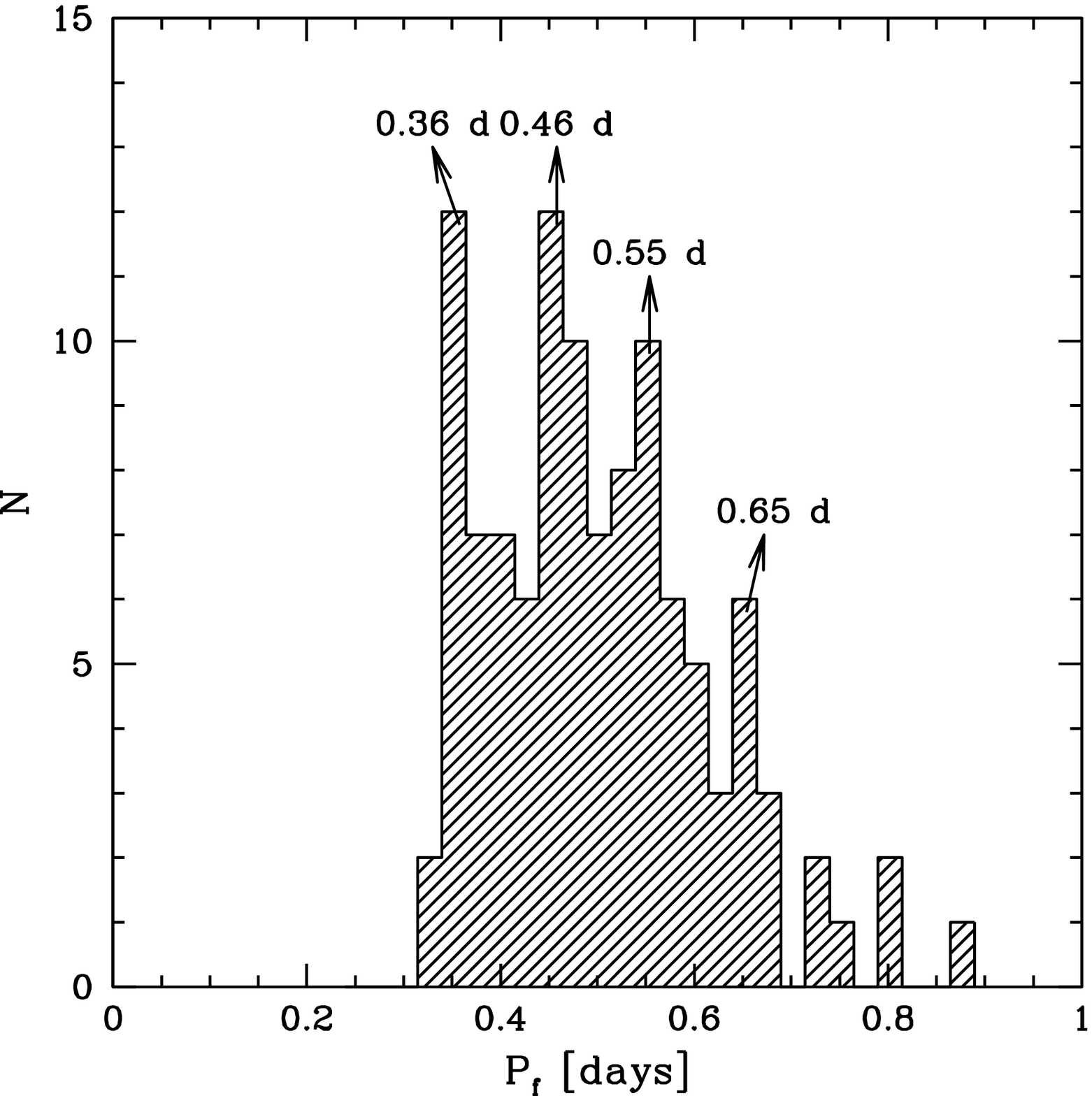}}
\caption{({\em top}) Period distribution of RR Lyrae stars in M14 from our work (solid line histogram) along with the distribution from the original sample of variable stars detected in WF94 (shaded histogram). The red, long-dashed, shaded histogram represents the area where RR Lyrae stars are considered to be potential second-overtone pulsators. ({\em bottom}) Fundamentalized period distribution of the RR Lyrae stars in M14. We note that V94, V108, V113 and V115, which are low-amplitude RRc stars according to CDLM12 but were not detected in our study, are not included in this plot.}
\label{fp_hist}
\end{center}
\end{figure}

Figure~\ref{fp_hist} (upper panel) shows the period distribution of RR Lyrae stars in M14, compared with the earlier results from WF94. We find a significantly larger fraction of short-period variables which may be attributed to ISIS's ability to find low-amplitude variables in crowded regions. In fact, 18/38 of the RRc stars detected with image subtraction techniques by CDLM12 and the present study are found within $1\arcmin$~ from the cluster center, with this number increasing to 26/38 within the innermost $2\arcmin$. Two main peaks are present, $P \sim 0.30$ and 0.55~d, characteristic of the presence of RRc and RRab stars, respectively. The short-period peak might tentatively be divided into two components, with periods peaking around $P \sim 0.28$ and 0.34~d. While the statistics is insufficient to establish the presence of these two peaks as significant, we nonetheless tentatively treat the shorter-period component as potentially being associated with second-overtone pulsators (RRe stars), given the presence of similar (and significant) such peaks in \citet{1996Alcock}. As can be seen from the figure, this putative shorter-period peak was also present in the results of WF94, but with a smaller number of stars. We thus tentatively classify as an RRe candidate any RR Lyrae variable with $P < 0.2897$~d. This period was defined as the limit of the bin containing all of the RR Lyrae that contributed to the tentative RRe stars' peak in the period distribution. We will further discuss the nature of these RRe candidates in Sect.~\ref{sec:rre} below, where it will be argued that these stars are in fact unlikely to correspond to a different RR Lyrae pulsation mode.   

In Figure~\ref{fp_hist} (bottom panel) we also show the distribution of fundamentalized periods\footnote{Periods were fundamentalized by adding 0.128 to their log-periods; see, e.g., \citet{2004Catelan}.} of the M14 RR Lyrae. Here we do not observe a sharply peaked distribution, contrary to what is seen in several other clusters, most notably M3 \citep[e.g.,][]{2004Catelan}. Four distinctive peaks are observed at $P_{f}\sim 0.36$, 0.46, 0.55, and 0.65~d. The observed feature at 0.55 days is also apparent in the GCs M15, M68, and M3 in the work of \citet{2004Catelan}, and here as in M3, it is primarily ascribed to RRab stars. The relatively large number of variables with periods $P_{f} < 0.5$~d is noteworthy; such a significant presence of short-period RRc stars is not observed in most of the clusters studied by \citet{2004Catelan} (the exception being M5). We also note the presence of a secondary hump at $P_{f}\sim0.65$~d, a characteristic that was also apparent in most of the GCs in \citet{2004Catelan}.

Lack of phase coverage for LPVs makes it hard to locate them precisely in the CMD. However, these variables are located close to the tip of the RGB, while type II Cepheids appear brighter than the HB, but with similar colors (Fig.~\ref{cmd1}). These are expected trends for these kinds of variable stars \citep[e.g.,][]{CS15}. Both the W~UMa and DEB stars are located above the HB in the CMD of Figure~\ref{cmd1}, and a more detailed discussion regarding these two stars is presented later in this paper.

\begin{deluxetable*}{lccccc}
\tabletypesize{\scriptsize} 
\tablecaption{Properties of variable stars found in M14.\tablenotemark{a}}
\tablehead{\colhead{ID} & \colhead{$\alpha$} & \colhead{$\delta$} & \colhead{$P$} & \colhead{$\Delta E(B-V)$} &\colhead{Type} \\
 & (J2000) & (J2000) & (d)  
}
\startdata
{V1}  &  {17:37:37.4}  &  {-03:13:59.5}  &  {18.76018}  &  {-0.006}  &  {Cepheid}\\
{V2}  &  {17:37:28.6}  &  {-03:16:45.2}  &  { 2.79471}  &  {-0.024}  &  {Cepheid}\\
{V3}  &  {17:37:36.1}  &  {-03:16:14.5}  &  { 0.52244}  &  {-0.001}  &  {RRab   }\\
{V4}  &  {17:37:47.0}  &  {-03:13:31.9}  &  { 0.65132}  &  {-0.001}  &  {RRab   }\\
{V5}  &  {17:37:27.2}  &  {-03:13:17.8}  &  { 0.54880}  &  {-0.016}  &  {RRab   }\\
{V6}  &  {17:37:38.6}  &  {-03:16: 2.3}  &  { -- }  &  {-0.028}  &  {LPV    }\\
{V7}  &  {17:37:40.4}  &  {-03:16:20.5}  &  {13.60380}  &  {-0.001}  &  {Cepheid}\\
{V8}  &  {17:37:42.6}  &  {-03:14: 9.5}  &  { 0.68607}  &  { 0.005}  &  {RRab   }\\
{V9}  &  {17:37:46.4}  &  {-03:15:23.8}  &  { 0.53881}  &  {-0.014}  &  {RRab   }\\
{V10}  &  {17:37:33.0}  &  {-03:18: 9.7}  &  { 0.58591}  &  {-0.007}  &  {RRab   }
\enddata
\tablenotetext{a}{See the text for a description of the columns. In here we only show a portion of the table for reference regarding its form and content. The entire table is available online.}
\label{varparsa}
\end{deluxetable*}

\begin{deluxetable*}{lcccccccccccccc}
\tabletypesize{\scriptsize} 
\tablecaption{Properties of variable stars found in M14.\tablenotemark{a}}
\tablehead{\colhead{ID} & \colhead{$A_{B}$} & \colhead{$A_{V}$} & \colhead{$A_{I}$} & \colhead{$(B)$} & \colhead{$(V)$} & \colhead{$(I)$} &  \colhead{$\langle B \rangle$} & \colhead{$\langle V \rangle$} & \colhead{$\langle I \rangle$} & \colhead{$(B-V)$} & \colhead{$\langle B\rangle-\langle V \rangle$} & \colhead{$(B-V)_{\mathrm{s}}$} & \colhead{$(V-I$)} & \colhead{$\langle V \rangle-\langle I \rangle$} }
\startdata
{V1}  &  {1.71}  &  {1.24}  &  {0.95}  &  {15.563}  &  {14.289}  &  {12.685}  &  {15.429}  &  {14.210}  &  {12.633}  &  {1.274}  &  {1.219}  &  { -- }  &  {1.605}  &  {1.577}\\
{V2}  &  {1.22}  &  {0.88}  &  {0.69}  &  {16.672}  &  {15.667}  &  {14.358}  &  {16.596}  &  {15.629}  &  {14.337}  &  {1.005}  &  {0.967}  &  { -- }  &  {1.309}  &  {1.292}\\
{V3}  &  {0.54}  &  {0.42}  &  {0.27}  &  {17.603}  &  {16.744}  &  {15.613}  &  {17.587}  &  {16.734}  &  {15.609}  &  {0.860}  &  {0.853}  &  {0.859}  &  {1.131}  &  {1.125}\\
{V4}  &  {1.24}  &  {0.93}  &  {0.58}  &  {18.148}  &  {17.211}  &  {15.977}  &  {18.079}  &  {17.177}  &  {15.964}  &  {0.937}  &  {0.902}  &  {0.927}  &  {1.234}  &  {1.213}\\
{V5}  &  {1.51}  &  {1.19}  &  {0.82}  &  {18.129}  &  {17.253}  &  {16.054}  &  {18.025}  &  {17.195}  &  {16.032}  &  {0.876}  &  {0.830}  &  {0.862}  &  {1.199}  &  {1.163}\\
{V6}  &  { -- }  &  { -- }  &  { -- }  &  {16.771}  &  {14.507}  &  {11.687}  &  { -- }  &  { -- }  &  { -- }  &  {2.265}  &  { -- }  &  { -- }  &  {2.819}  &  { -- }\\
{V7}  &  {1.28}  &  {0.88}  &  {0.70}  &  {16.123}  &  {14.786}  &  {13.251}  &  {16.051}  &  {14.745}  &  {13.224}  &  {1.337}  &  {1.306}  &  { -- }  &  {1.535}  &  {1.521}\\
{V8}  &  {0.76}  &  {0.54}  &  {0.36}  &  {18.194}  &  {17.225}  &  {15.932}  &  {18.168}  &  {17.213}  &  {15.927}  &  {0.968}  &  {0.955}  &  {0.958}  &  {1.294}  &  {1.286}\\
{V9}  &  {1.39}  &  {1.06}  &  {0.70}  &  {18.254}  &  {17.314}  &  {16.096}  &  {18.162}  &  {17.262}  &  {16.075}  &  {0.939}  &  {0.900}  &  {0.928}  &  {1.218}  &  {1.187}\\
{V10}  &  {1.35}  &  {1.10}  &  {0.71}  &  {17.954}  &  {17.186}  &  {16.017}  &  {17.878}  &  {17.137}  &  {15.999}  &  {0.768}  &  {0.741}  &  {0.757}  &  {1.169}  &  {1.138}
\enddata
\tablenotetext{a}{See the text for a description of the columns. In here we only show a portion of the table for reference regarding its form and content. The entire table is available online.}
\label{varparsb}
\end{deluxetable*}

\section{Fourier Decomposition}\label{sec_fourier}

Fourier decomposition consists in fitting the light curves of variable stars as a sum of sine and/or cosine functions. This method allows us to better constrain the mean color and magnitude of the star, compared to the results given by ALLFRAME, and to derive a set of physical parameters for each star.

In our analysis, the $V$-band light curves of pulsating stars were fitted to an equation of the form \citep[following][]{corwin_m75}

\begin{equation}
mag=A_{0}+\sum_{j=1}^{N}{A_{j}\sin(j\omega t+\phi_{j}+\Phi)},
\end{equation}  

\noindent where $\omega=2\pi/P$, and the light curve shape is quantified in terms of the lower-order coefficients $A_{j1}$=$A_{j}/A_{1}$ and $\phi_{j1}=\phi_{j}-j\phi_{1}$. For RRab stars we use $\Phi=0$, whilst for RRc objects, $\Phi=\pi/2$. In order to avoid unrealistic fluctuations in the fits, the order of the Fourier series was tweaked on a star-by-star basis, upon visual inspection of the results; however, in most cases, $N>6$.

\citet{sicle_rrc} demonstrated that Fourier decomposition can be very useful to obtain physical parameters of RR Lyrae stars. They used hydrodynamical simulations along with observations of RRc stars in several GCs to propose a calibration of their physical parameters (including luminosity and mass) as a function solely of two observational parameters, namely the period and $\phi_{31}$. Using equations (2), (3), (6), and (7) from \citet{sicle_rrc}, one can derive the mass, luminosity, and effective temperature of the star, as well as a ``helium parameter'' $Y$. Great care has to be taken when analyzing the results for this last parameter, as it does not necessarily correspond to actual He abundances in the ionization zones nor to accurate main-sequence helium abundances \citep{sicle_rrc}. Note, in addition, that, as pointed out by \citet{cat_rrc1}, the equations for mass and luminosity cannot be simultaneously valid, since by combining them one arrives at an expression that is formally inconsistent with Ritter's period-mean density relation \citep[see also][]{2010Deb}. Given these caveats, the masses, luminosities, effective temperatures and helium abundances of RRc stars, as derived using the Fourier decomposition method, will not be reported in this paper. However, the interested reader may easily obtain these parameters, using the data provided in Tables~\ref{varparsa} and \ref{foparamrrc} and the aforementioned equations from \citet{sicle_rrc}.\footnote{In Table~\ref{tab_corwin2} below, the resulting averages over the cluster variables are listed, though solely with the purpose of providing a comparison with other GCs for which the same quantities have been similarly derived in the literature.}

Based on Fourier composition, we can also obtain two additional parameters for RRc stars, the intensity-averaged absolute magnitude $M_{V}$ \citep[from][]{kov1998_2} and the metallicity ${\rm [Fe/H]}$ \citep[from][]{morgan2007}, using the following:

\begin{eqnarray}
{\rm [Fe/H]} & = & 0.367 - 8.507 \, P + 0.196 \, \phi_{31} + 0.0348 \, \phi^{2}_{31},\label{rrc1}\\
\langle M_{v}\rangle & = & 1.061- 0.961 \, P - 4.447 \, A_{4} + 0.044 \, \phi_{21}.\label{rrc2}
\end{eqnarray}

\noindent We note that the zero point in equation~\ref{rrc2} is 0.2 mag brighter than the original value presented in \citet{kov1998_2}, in order to make the luminosities of the RRc consistent with the distance modulus of 18.5 mag for the LMC \citep[following][]{2010Arellano}. Finally, ${\rm [Fe/H]}$ is in the metallicity scale of \citet{cg1}; \citet{morgan2007} also present an empirical equation in the \citet{zw84} metallicity scale (their eq.~3). In this work, we transform both values to the UVES \citet{cg2} metallicity scale, using the transformations provided in \citet{cg2}. Finally, the metallicity for RRc stars is given by the mean between these two values. 

\citet{jur_kov96}, \citet{kov_jur96,kov_jur97}, \citet{jur98}, \citet{kov_kan98}, and \citet{kov_wal99,kov_wal01} derive empirical equations that relate the metallicity, absolute magnitude, intrinsic colours and temperature of a star to Fourier decomposition parameters and the observed fundamental period, $P_{0}$. However, it has been noted that the values obtained for intrinsic colours show significant discrepancies with respect to the observed values, and are thus unreliable \citep[see e.g.][]{cac2005}. For this reason, these intrinsic colors, and accordingly the effective temperatures derived therefrom, are not reported in our paper, though again the interested reader may easily compute them from the expressions in the aforementioned papers and the periods and Fourier coefficients given in Tables~\ref{varparsa} and \ref{foparamab}. The Fourier-based parameters that we do report are the following:

\begin{eqnarray}
{\rm [Fe/H]}_{\rm J95} & = & -5.038 - 5.394 \, P + 1.345 \, \phi_{31},\label{jk1}\\
\langle M_{v}\rangle & = & -1.867 \, \log P - 1.158 \, A_{1} + 0.821\, A_{3} + K,\label{jk2}
\end{eqnarray}

\noindent
where the zero point for equation~\ref{jk2}, $K$, is adopted as $K=0.41$~mag \citep[see][]{2010Arellano}. These equations are valid for stars satisfying the ``compatibility condition,'' i.e., with deviation parameters $D_{m} \leq 3$. The $D_{m}$ parameter was introduced by \citet{jur_kov96} as a test of the precision of these equations and is defined as the maximum of 9 deviation parameters, $\{D_{F}\}$, defined as

\begin{align}
D_{F}=\frac{|F_{\rm obs}-F_{jk}|}{\sigma_{F}}, 
\label{eq:devpar}
\end{align}

\noindent
where $F_{\rm obs}$ is the observed value of the given Fourier parameter, $F_{jk}$ is its predicted value using the other observed parameters, and $\sigma_{F}$ is the standard deviation of $F_{jk}$ \citep{jur_kov96}. The expressions for $F_{jk}$ as well as the values of $\sigma_{F}$ can be found in Table~6 of \citet{jur_kov96}. \citet{jur_kov96} argue that light curves showing $D_{m} \leq 3$ satisfy the \textit{compatibility condition} and physical parameters derived from these stars can be accounted as reliable estimates of such parameters. 

Tables~\ref{foparamrrc} and \ref{foparamab} show the amplitude ratios and phase differences obtained for the RRc and RRab stars, respectively. In these tables, a ``:'' marks stars with uncertain values, which we do not use in the subsequent analysis. Stars whose amplitude ratios and phase differences are deemed particularly unreliable are omitted from these tables. 

\begin{deluxetable}{lcccccc}
\tablecaption{Amplitude ratios and phase differences from Fourier analysis for c-type RR Lyrae stars} 
\tablehead{
\colhead{ID} & \colhead{$A_{21}$} &  \colhead{$A_{31}$}  & \colhead{$A_{41}$}  & \colhead{$\phi_{21}$}  & \colhead{$\phi_{31}$}  & \colhead{$\phi_{41}$} 
}
\startdata
V20 & 0.201 & 0.072 & 0.068 & 4.637 & 2.433 & 1.437 \\
V21 & 0.082 & 0.066 & 0.061 & 4.665 & 3.927 & 2.278 \\
V25: & 0.098 & 0.044 & 0.076 & 4.991 & 4.410 & 3.217 \\
V44: & 0.136 & 0.108 & 0.098 & 4.663 & 3.096 & 2.005 \\
V46: & 0.079 & 0.060 & 0.052 & 4.716 & 5.557 & 2.350 \\
V51: & 0.134 & 0.106 & 0.015 & 3.630 & 2.416 & 1.780 \\
V55 & 0.069 & 0.086 & 0.051 & 4.912 & 4.207 & 2.325 \\
V56 & 0.069 & 0.083 & 0.089 & 5.677 & 3.816 & 2.074 \\
V58: & 0.038 & 0.137 & 0.095 & 3.986 & 4.128 & 2.805 \\
V78 & 0.098 & 0.070 & 0.017 & 4.657 & 3.776 & 3.079 \\
V80 & 0.103 & 0.044 & 0.047 & 5.552 & 3.755 & 2.628 \\
V88: & 0.096 & 0.066 & 0.068 & 5.048 & 3.829 & 2.981 \\
V90: & 0.062 & 0.055 & 0.043 & 4.369 & 5.484 & 2.156 \\
V91: & 0.135 & 0.034 & 0.011 & 4.428 & 2.685 & 1.951 \\
V95 & 0.079 & 0.072 & 0.030 & 4.662 & 3.868 & 3.324 \\
V96 & 0.142 & 0.030 & 0.011 & 4.920 & 3.831 & 0.040 \\
V98: & 0.177 & 0.054 & 0.021 & 4.619 & 3.261 & 3.220 \\
V104: & 0.196 & 0.058 & 0.041 & 4.506 & 4.123 & 0.828 \\
V105 & 0.157 & 0.089 & 0.073 & 4.393 & 2.943 & 1.389 \\
V110 & 0.057 & 0.074 & 0.047 & 4.966 & 2.956 & 1.787 \\
V117: & 0.040 & 0.091 & 0.094 & 6.174 & 4.483 & 1.651 \\
V131: & 0.149 & 0.112 & 0.040 & 4.552 & 2.533 & 1.255 \\
V133 & 0.127 & 0.082 & 0.062 & 4.678 & 3.478 & 2.276 \\
V135 & 0.073 & 0.096 & 0.035 & 4.763 & 4.354 & 2.908 \\
V136 & 0.127 & 0.116 & 0.040 & 4.983 & 4.239 & 1.633 \\
V138: & 0.043 & 0.134 & 0.038 & 3.845 & 3.927 & 2.677 \\
V143: & 0.060 & 0.102 & 0.027 & 5.363 & 4.146 & 3.942 \\
V148: & 0.120 & 0.023 & 0.041 & 4.840 & 1.880 & 6.207 \\
V153: & 0.071 & 0.059 & 0.062 & 3.792 & 0.079 & 6.093 \\
V154 & 0.128 & 0.036 & 0.028 & 4.624 & 2.274 & 2.312 \\
V157 & 0.181 & 0.059 & 0.061 & 4.685 & 3.062 & 1.928 \\
V159: & 0.182 & 0.076 & 0.044 & 4.734 & 2.986 & 2.165 \\
V160: & 0.108 & 0.093 & 0.012 & 5.027 & 3.418 & 1.212 \\
V163 & 0.152 & 0.101 & 0.049 & 4.512 & 2.858 & 1.734 
\enddata
\label{foparamrrc}
\end{deluxetable} 

In Table~\ref{fophys_rrc} we show the values of the metallicity (in the UVES scale) and absolute magnitudes of RRc stars obtained from Fourier decomposition. We note that the mean metallicity, ${\rm [Fe/H]}_{\rm UVES} = -1.20 \pm 0.27$~dex, is consistent with the listed value in the \citet{h96} catalog and our results from the analysis of the CMD of M14 \citep{2013Contreras}. Note that the revised [Fe/H] relation from \citet{2014Morgan} leads to similar results, i.e., ${\rm [Fe/H]}_{\rm UVES}=-1.22\pm0.29$~dex.

\begin{deluxetable}{lccccccc}
\tablecaption{Amplitude ratios and phase differences for RRab stars\tablenotemark{a}} 
\tablehead{
\colhead{ID} & \colhead{$A_{21}$} &  \colhead{$A_{31}$}  & \colhead{$A_{41}$}  & \colhead{$\phi_{21}$}  & \colhead{$\phi_{31}$}  & \colhead{$\phi_{41}$} & \colhead{$D_{m}$} 
}
\startdata
V4 & 0.556 & 0.340 & 0.152 & 2.675 & 5.340 & 2.137 & 5.3   \\
V5 & 0.525 & 0.342 & 0.245 & 2.362 & 5.131 & 1.565 & 1.4   \\
V8: & 0.471 & 0.285 & 0.093 & 2.707 & 5.902 & 2.861 & 8.7   \\
V9 & 0.451 & 0.250 & 0.158 & 2.385 & 5.163 & 1.224 & 6.7   \\
V10: & 0.523 & 0.310 & 0.197 & 2.415 & 5.176 & 1.671 & 2.4   \\
V11 & 0.477 & 0.344 & 0.255 & 2.311 & 4.931 & 1.249 & 2.9   \\
V13: & 0.528 & 0.312 & 0.132 & 2.343 & 5.020 & 2.050 & 6.2   \\
V14: & 0.539 & 0.365 & 0.227 & 2.532 & 5.251 & 1.416 & 8.2   \\
V15 & 0.547 & 0.317 & 0.257 & 2.479 & 5.154 & 1.735 & 4.7   \\
V16: & 0.575 & 0.339 & 0.220 & 2.418 & 5.021 & 1.700 & 6.4   \\
V18 & 0.455 & 0.321 & 0.167 & 2.377 & 4.920 & 1.358 & 4.0   \\
V19 & 0.543 & 0.309 & 0.232 & 2.411 & 5.180 & 1.588 & 3.0   \\
V22: & 0.532 & 0.335 & 0.155 & 2.613 & 5.381 & 2.278 & 8.1   \\
V23 & 0.427 & 0.250 & 0.134 & 2.443 & 4.974 & 1.267 & 39.8  \\
V24 & 0.549 & 0.354 & 0.306 & 2.151 & 4.848 & 1.244 & 39.0  \\
V30 & 0.549 & 0.359 & 0.221 & 2.276 & 5.038 & 1.222 & 5.0   \\
V31 & 0.508 & 0.307 & 0.219 & 2.466 & 5.069 & 1.721 & 6.0   \\
V32 & 0.461 & 0.273 & 0.118 & 2.717 & 5.482 & 2.318 & 11.7  \\
V33 & 0.482 & 0.297 & 0.215 & 2.173 & 4.639 & 0.822 & 48.7  \\
V34 & 0.557 & 0.287 & 0.175 & 2.440 & 5.244 & 1.918 & 3.6   \\
V36: & 0.566 & 0.311 & 0.077 & 2.640 & 5.725 & 2.677 & 7.8   \\
V37 & 0.512 & 0.423 & 0.237 & 2.244 & 4.989 & 1.070 & 8.7   \\
V38: & 0.457 & 0.318 & 0.230 & 2.331 & 4.867 & 1.509 & 4.1   \\
V39 & 0.610 & 0.325 & 0.304 & 2.313 & 5.029 & 1.472 & 5.9   \\
V42: & 0.501 & 0.341 & 0.157 & 2.389 & 5.048 & 1.450 & 4.1   \\
V43 & 0.443 & 0.326 & 0.227 & 2.230 & 4.754 & 1.052 & 40.5  \\
V48 & 0.496 & 0.267 & 0.182 & 2.347 & 5.058 & 1.162 & 7.0   \\
V49: & 0.477 & 0.240 & 0.087 & 2.678 & 5.604 & 2.269 & 8.0   \\
V57 & 0.602 & 0.208 & 0.083 & 2.294 & 5.341 & 1.767 & 85.5  \\
V60 & 0.484 & 0.359 & 0.143 & 2.662 & 5.415 & 2.156 & 5.9   \\
V61 & 0.470 & 0.381 & 0.206 & 2.426 & 5.186 & 1.654 & 9.0   \\
V62 & 0.340 & 0.150 & 0.039 & 2.549 & 5.681 & 4.808 & 143.3 \\
V68: & 0.476 & 0.346 & 0.205 & 2.323 & 5.006 & 1.393 & 5.8   \\
V70 & 0.488 & 0.268 & 0.196 & 2.515 & 5.207 & 1.777 & 7.7   \\
V71 & 0.423 & 0.148 & 0.094 & 2.270 & 4.638 & 0.938 & 19.6  \\
V75: & 0.489 & 0.291 & 0.256 & 2.369 & 5.020 & 1.451 & 7.9   \\
V77 & 0.386 & 0.141 & 0.102 & 2.965 & 5.975 & 3.390 & 115.4 \\
V79 & 0.512 & 0.234 & 0.259 & 2.715 & 5.389 & 1.830 & 10.3  \\
V92 & 0.077 & 0.121 & 0.132 & 3.614 & 4.004 & 4.338 & 112.4 \\
V111: & 0.458 & 0.272 & 0.103 & 3.128 & 6.203 & 4.336 & 138.2 \\
V122 & 0.612 & 0.402 & 0.267 & 2.088 & 4.990 & 1.354 & 5.3   \\
V124: & 0.455 & 0.271 & 0.156 & 2.532 & 5.279 & 1.686 & 7.3   \\
V130: & 0.441 & 0.224 & 0.134 & 2.464 & 5.232 & 1.969 & 20.9  \\
V132 & 0.569 & 0.387 & 0.352 & 2.539 & 5.274 & 1.430 & 5.4   \\
V140 & 0.430 & 0.166 & 0.051 & 2.825 & 6.049 & 3.680 & 122.6 \\
V141 & 0.509 & 0.333 & 0.195 & 2.554 & 5.435 & 2.316 & 6.3   \\
V142 & 0.485 & 0.279 & 0.292 & 2.174 & 4.557 & 0.828 & 47.8  \\
V147 & 0.531 & 0.362 & 0.314 & 2.334 & 4.680 & 1.133 & 18.5  \\
V150 & 0.451 & 0.205 & 0.086 & 3.037 & 6.097 & 3.584 & 123.4 \\
V166: & 0.387 & 0.176 & 0.044 & 2.167 & 4.725 & 2.551 & 50.3  
\enddata
\tablenotetext{a}{The last column represents the $D_{m}$ parameter (see main text).}
\label{foparamab}
\end{deluxetable}

\begin{deluxetable}{lcc}
\tablecaption{Derived parameters for c-type RR Lyrae stars} 
\tablehead{
\colhead{ID} & \colhead{${\rm [Fe/H]}_{\rm UVES}$}  & \colhead{$\langle M_{V} \rangle$}  
}
\startdata
V20 & -1.317$\pm$0.114 & 0.531$\pm$0.023  \\
V21 & -1.139$\pm$0.139 & 0.486$\pm$0.023  \\
V55 & -1.177$\pm$ 0.129 & 0.473$\pm$0.025   \\
V78 & -1.131$\pm$0.088 & 0.542$\pm$0.020   \\
V80 & -1.207$\pm$0.244 & 0.471$\pm$0.025   \\
V95 & -1.627$\pm$0.107 & 0.483$\pm$0.020   \\
V96 & -0.523$\pm$0.314 & 0.593$\pm$0.018  \\
V110 & -1.492$\pm$0.137 & 0.508$\pm$0.023  \\
V133 & -1.221$\pm$0.150 & 0.501$\pm$0.021  \\
V135 & -1.098$\pm$0.132 & 0.492$\pm$0.022 \\
V154 & -1.293$\pm$0.486 & 0.589$\pm$0.020  \\
V157 & -0.984$\pm$0.138 & 0.535$\pm$0.020   \\
V163 & -1.417$\pm$0.091 & 0.529$\pm$0.020  \\ 
\hline 
Mean  & -1.202$\pm$0.268 & 0.518$\pm$0.040  
\enddata
\label{fophys_rrc}
\end{deluxetable}

In Table~\ref{foparamab}, the last column stands for the ``deviation parameter'' $D_{m}$ value (see eq.~\ref{eq:devpar} and the corresponding discussion). Using equations~\ref{jk1} and \ref{jk2} we derive physical parameters for stars with $D_{m} <  5$.\footnote{As previously stated, the original criterion of \citet{jur_kov96} for differentiating ``normal''  RRab from those with inadequate light curves for their method to work is $D_{m} < 3$. However, following \citep{2010Contreras}, and in order to get a larger final sample, we have opted to relax this constraint slightly.} The metallicities and absolute magnitudes for stars which follow this condition are presented in Table~\ref{fophys_rrab}; the values in this table are derived on the basis of equations~\ref{jk1}-\ref{jk2}. 

\begin{deluxetable}{lcc}
\tablecaption{Physical parameters of ab-type RR Lyrae\tablenotemark{a}}
\tablehead{
\colhead{ID} & \colhead{${\rm [Fe/H]}_{\rm J95}$}  & \colhead{$\langle M_{V} \rangle$} 
}
\startdata
V5 & -1.097$\pm$0.091 & 0.551$\pm$0.014  \\
V10* & -1.237$\pm$0.123 & 0.506$\pm$0.016 \\
V11* & -1.666$\pm$0.178 & 0.467$\pm$0.018  \\
V15 & -1.115$\pm$0.105& 0.529$\pm$0.016  \\
V18 & -1.005$\pm$0.062 & 0.661$\pm$0.013  \\
V19 & -1.014$\pm$0.079 & 0.539$\pm$0.014  \\
V30 & -1.143$\pm$0.093 & 0.575$\pm$0.014  \\
V34 & -1.257$\pm$0.208 & 0.574$\pm$0.016  \\
V38* & -1.234$\pm$0.079 & 0.618$\pm$0.014 \\
V42* & -1.652$\pm$0.135 & 0.579$\pm$0.015  \\
V59* & -0.994$\pm$0.173 & 0.592$\pm$0.015 \\
V158* & -1.064$\pm$3.079 & 0.574$\pm$0.013  \\
\hline
Mean\tablenotemark{b} & -1.105$\pm$0.093 &  0.572$\pm$0.048  

\enddata
\tablenotetext{a}{Derived from Fourier parameters of stars with $D_{m} < 5$.}
\tablenotetext{b}{Mean values are derived excluding variable stars with uncertain parameters (marked by an asterisk) as explained in the main text.}
\label{fophys_rrab}
\end{deluxetable}

In Table~\ref{fophys_rrab}, V11 stands out as being the intrinsically brightest of M14's RR Lyrae, at least among those satisfying $D_{m} < 5$. In the work of WF94, this variable was already noted to be overluminous, compared to the rest of the RR Lyrae stars in M14, and the same is noted in our data. The star does not present clear signs of a blend with an unresolved companion. Variable stars V10, V38, V42, V59, and V158 have uncertain Fourier parameters, and thus will not be considered in the following analysis. By not considering these variable stars we find ${\rm [Fe/H]}_{\rm J95}=-1.11\pm0.09$~dex in the \citet{Jur1995} metallicity scale, which is transformed to the \citet{zw84} scale using

\begin{equation}
{\rm [Fe/H]}_{\rm J95} = 1.431 \, {\rm [Fe/H]}_{\rm ZW}+0.880. 
\end{equation}

\noindent
This yields a metallicity for M14 of ${\rm [Fe/H]}_{\rm ZW84}=-1.39\pm0.09$~dex, or ${\rm [Fe/H]}_{\rm UVES}=-1.28\pm0.09$~dex in the most recent UVES metallicity scale of \citeauthor{cg2}, based on the quadratic transformation between ${\rm [Fe/H]}_{\rm ZW84}$ and ${\rm [Fe/H]}_{\rm UVES}$ provided in \citet{cg2}. This is close to the listed value for this GC and in excellent agreement with the derived value from the analysis of the CMD of M14 \citep[see][]{2013Contreras}. It is also consistent with the value derived above from analysis of the RRc light curves, to within the errors.\footnote{Prompted by the referee, we have also obtained [Fe/H] values for the M14 variables using the expressions derived by \citet{2013Nemec}. Based on their eqs.~(2) and (3) alike, we find ${\rm [Fe/H]} = -1.1$~dex. This value is significantly higher than reported above, based on the original \citet{cg2} transformations, and also than found in our Paper~I (based on the cluster CMD) or provided in the \citet{h96,wh10} catalog. In addition, the fact that both of the aforementioned \citet{2013Nemec} equations provide the same [Fe/H] value is unexpected, as their eq.~(3) in particular should provide [Fe/H] values in the \citet{Jur1995} system, which is different from the UVES one. The reason for this inconsistency is unclear, and explaining it is beyond the scope of our paper.}

\section{Distance Modulus}\label{sec:DM} 

We can use the results from the previous section to estimate the distance modulus of M14. By using the mean absolute magnitude, $\langle M_{V} \rangle = 0.572\pm0.048$~mag, and the mean magnitude, $\langle V \rangle_{\rm RRab} = 17.19\pm0.04$~mag, of the RRab stars in Table~\ref{fophys_rrab} (derived from the intensity-weighted values of Table~\ref{varparsb}, without considering stars marked by an asterisk), we obtain a distance modulus $(m-M)_{V} = 16.61\pm0.06$~mag. Similarly, for the RRc stars in Table~\ref{fophys_rrc}, with $\langle V \rangle_{\rm RRc} = 17.24\pm0.11$~mag and $\langle M _{V} \rangle = 0.518\pm0.040$~mag, one finds $(m-M)_{V} = 16.72\pm0.12$~mag. While these two estimates are in agreement to within the errors, it should be noted that the nominal difference between them is due in large part to the fainter zero point adopted in the RRab $M_{V}$ calibration. A weighted average of the two results gives a distance modulus of $(m-M)_V = 16.63 \pm 0.13$~mag, which is in good agreement with the value provided in the \citet{h96,wh10} catalog. Adopting a standard extinction law with $A_{V}=3.1\,E(B-V)$ and our estimated reddening of $E(B-V)=0.57\pm0.02$~mag \citep{2013Contreras}, we get a true distance modulus of $(m-M)_0=14.86\pm0.13$~mag. If instead we use the $M_V-{\rm [Fe/H]}$ relation from \citet{2008Catelan_cortes}, we obtain $\langle M_{V} \rangle = 0.66\pm0.14$~mag, implying a true distance modulus of $(m-M)_0=14.77\pm0.16$~mag.

\begin{figure}
\centering
\resizebox{\columnwidth}{!}{\includegraphics{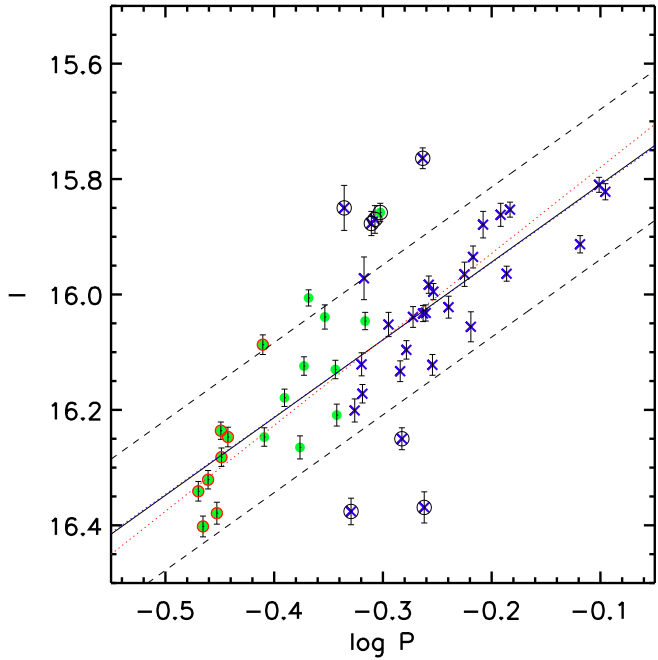}}
\caption{Period-luminosity relation for RR Lyrae stars in M14. RRab, RRc, and (candidate) RRe stars are shown with the same symbols as in Figure~\ref{cmd1}. The periods of the RRc and candidate RRe stars were fundamentalized as explained in the text. The solid line represents the period-luminosity relation from the theoretical calibration of \citet{cat_pir1} for a metallicity $Z = 0.001$, shifted by $(m-M)_{I}=15.90$~mag. Long-dashed lines at $\pm0.13$ mag show the same line shifted to match the upper and lower ends of the observed distribution. The dotted lines show the least-squares fits to the data assuming $\log P$ as the independent variable ({\em blue}) and using the OLS bisector expression in Table~1 of \citet[][{\em red}]{1990Isobe}. The stars marked by black circles were not included in the fits.} 
\label{pl}
\end{figure}

The distance modulus can also be obtained using the $I$-band period-luminosity relation \citep{cat_pir1}. To obtain a more precise relation, we restricted our analysis to those light curves for which high-order Fourier fits (i.e., fits with six or more terms) could be obtained, and the $\langle I \rangle$ values derived accordingly. As before, the periods of the RRc stars were fundamentalized by adding 0.128 to their values of $\log P$. The same procedure was adopted in the case of the RRe candidates. The result is shown in Figure~\ref{pl}, where the data are compared with the corresponding theoretical calibrations from \citet{cat_pir1} for a metallicity $Z=0.001$. This line was then vertically shifted to match our observed relation, giving an apparent distance modulus of $(m-M)_{I}=15.90\pm0.13$~mag, with the error conservatively estimated by shifting the line to match the upper and lower ends of the $I$-band distribution. In the same figure we compare the theoretical relation with two least-squares fit to the data, one using $\log P$ as the independent variable, and another using the ``OLS bisector'' expression from Table~1 in \citet{1990Isobe}.  In both fits we have not included the outliers highlighted with circles in Figure~\ref{pl}. The slope of the fit where we assume $\log P$ as the independent variable is remarkably similar to the slope of the theoretical relation. In the case of the OLS bisector fit, we find a slightly steeper slope, as can be seen in the figure. From the apparent distance modulus derived from the theoretical relation, with $E(B-V)=0.57\pm0.02$~mag \citep{2013Contreras} and assuming $A_{I}=1.48\, E(B-V)$ \citep{1989Cardelli}, we derive an intrinsic distance modulus of $(m-M)_0=15.05\pm0.14$~mag, which is similar (to within the errors) to the values obtained above. Thus, depending on the technique used and zero point adopted in the calibrations, one finds an RR Lyrae-based distance to M14 in the range between about 9.0 and 10.2~kpc.

\section{On the Nature of the RRe Stars}\label{sec:rre} 

Since the process of fundamentalization that we applied in the previous section is strictly applicable only to RRc stars, one would not expect second-overtone stars to follow the same linear trend as RRab and RRc stars in Figure~\ref{pl}; in fact, RRe candidates should be located systematically to the left of the theoretical relation: based on Table~5.2 in \citet{CS15}, one would rather expect that the offset between the RRab and RRe trends should amount to $\delta \log P \approx 0.25$, which is about twice as large as the value (0.128) adopted to fundamentalize the RRc and RRe stars' periods in Figure~\ref{pl}. However, Figure~\ref{pl} shows that our RRe candidates tend to conform rather well to the linear trend when an offset of 0.128 is applied, similarly to what had previously been found for other GCs \citep{cat_rrc1}. 

When the positions of the RRe candidates are compared against those of the longer-period RRc stars in the Bailey diagram, one finds that most of these candidates tend to show smaller pulsation amplitudes. However, we do note that a bell-shaped distribution in the Bailey diagram, with the shortest-period RRcs having the smallest amplitudes, is naturally predicted for RRc stars \citep[see, e.g., Figs.~12, 15 in][]{1997Bono}. In like vein, while the RRe candidates are located close to the blue edge of the instability strip (Fig.~\ref{cmd1}), which is where such second-overtone pulsators would likely be located if they really existed \citep[e.g.,][]{1996Walker}, short-period, low-amplitude RRc stars are also expected to inhabit the proximity of the instability strip blue edge. 

\begin{figure}
\centering
\resizebox{\columnwidth}{!}{\includegraphics{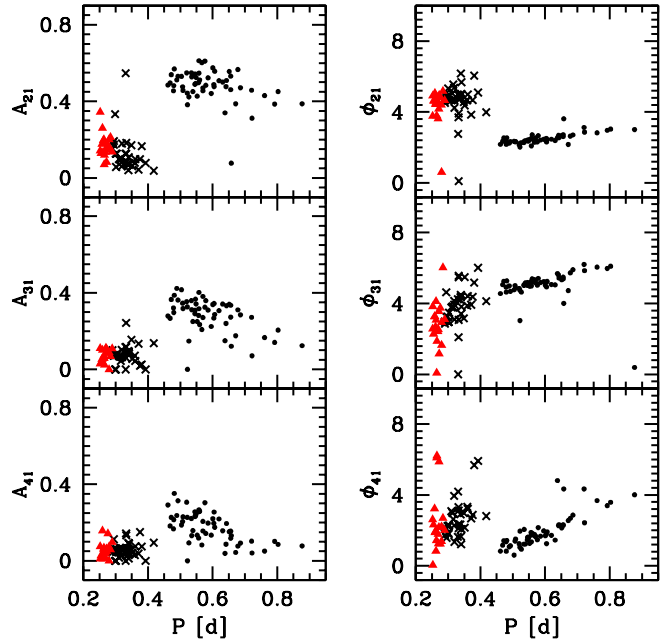}}
\caption{(Comparison between amplitude ratios and phase differences of RRab (\textit{solid circles}), RRc (\textit{crosses}) and candidate RRe stars (\textit{red triangles}). }
\label{fourier_rre}
\end{figure}

\citet{kovacs1998} argues that any difference in the morphology of the light curves of RRe stars, compared to first-overtone pulsators, would result in a difference in the Fourier decomposition parameters as a function of their periods, as is the case between RRab and RRc stars (especially in $A_{21}$ and $\phi_{21}$). In Figure~\ref{fourier_rre} we plot the results of our Fourier parameters for RRc and candidate RRe stars. We can see that there is no clear distinction between RRc stars and RRe candidates, similar to what was found by \citet{kovacs1998}.

The above arguments all strongly suggest that the candidate RRe stars in M14 do not reflect a different pulsation mode, but constitute simply the short-period end of the RRc distribution.

\section{The Oosterhoff Type of M14}\label{sec:oost}

WF94 classified M14 as an OoI GC. Their result is based on the mean period of the ab-type RR Lyrae and the relative fraction of RRc variables,  $f_{c} \equiv N_{c}/(N_{c}+N_{ab})$. They found $\langle P_{ab} \rangle = 0.572 $~d, which places M14 at the high end of the OoI clusters' period distribution.  However, when they exclude V47, arguing that the star shows an unusually long period and low amplitude compared to other RR Lyrae in their survey, the mean period drops to $\langle P_{ab} \rangle = 0.564$~d, closer to the mean value for OoI clusters \citep[see, e.g.,][]{1995Smith,CS15}. For the RRc number fraction, they obtain $f_{c}=0.273$ , which is also a typical value for OoI clusters.

We find 56 ab- and 54 c-type RR Lyrae in M14. If we assume that all of them are cluster members, we derive  $\langle P_{ab} \rangle = 0.589$~d, $\langle P_{c} \rangle = 0.309$~d, and $f_{c} = 0.491$. The value of $f_{c}$ is much higher than the previous estimate of WF94, and approaches the expected value for OoII clusters, which often contain a higher proportion of c-type RR Lyrae, compared to OoI globulars. However, this may be just an indication of the success of our study in finding low-amplitude variables, rather than an accurate indication of the intrinsic Oosterhoff status of M14; in fact, other clusters  \citep[e.g., M62, NGC\,6171, NGC\,4147;][]{rcon1,1997Clement,2004Arellano} present a high value of $f_{c}$, despite their OoI classification, thus suggesting that $f_{c}$ is not a particularly reliable indicator of Oosterhoff type. 

\begin{figure}
\centering
\resizebox{\columnwidth}{!}{\includegraphics{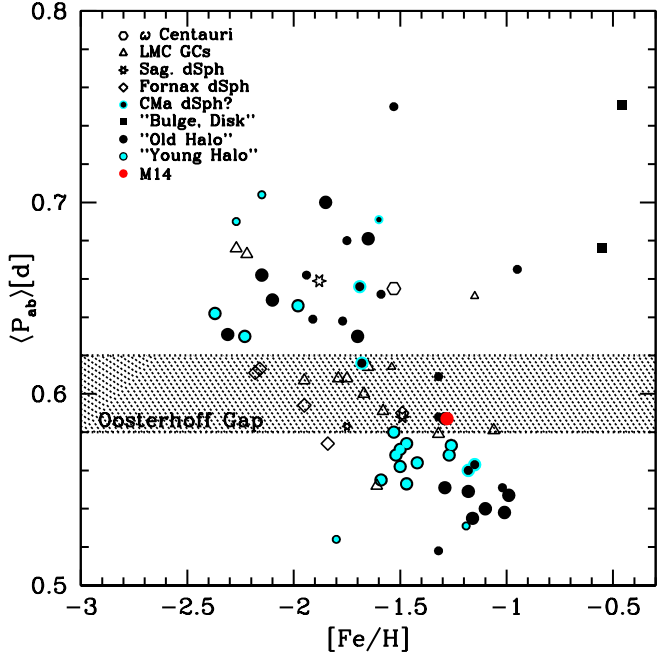}}
\caption{Location of M14 in the ${\rm [Fe/H]}-\langle P_{ab} \rangle$. }
 \label{osdich_all}
\end{figure}
 
Our derived $\langle P_{ab} \rangle$ value is also higher than the one from WF94. The four long-period RRab stars ($P>0.7$~d), V140, V158, V150, and V172, and the new period and classification for V77, make V47 less unusual; in fact, these variable stars also share similar light curve amplitudes with V47, and so we find no reason for excluding this variable from our analysis. $\langle {P}_{ab} \rangle$ is higher than would be expected for OoI GCs; our new results actually locate M14 inside the Oosterhoff gap (see Fig.~\ref{osdich_all}).

\begin{figure}
\begin{center}
 \resizebox{\columnwidth}{!}{\includegraphics{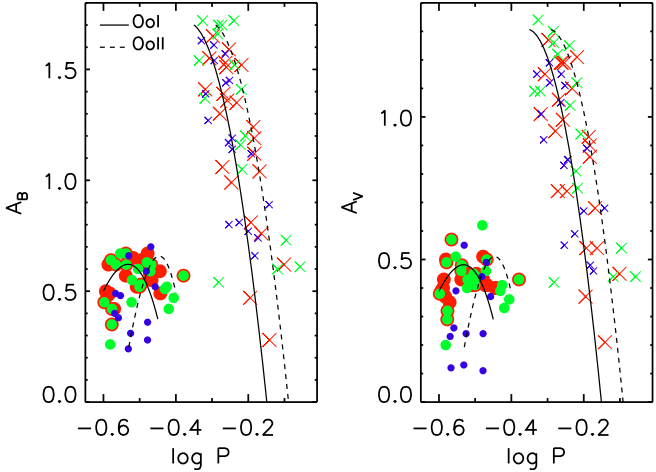}}
 \caption{$B$- ({\em left}) and $V$-band ({\em right}) Bailey diagrams for all of the RR Lyraes found in M14. RRab and RRc stars are shown as crosses and filled circles, respectively. Large red symbols mark stars which are located at a distance from the center of the cluster $r > 2 \, r_{c} $, where $r_{c} = 0.79\arcmin$ is the cluster's core radius. Medium-size green symbols represent stars with $2 \, r_{c} \geq r \geq r_{c}$, while small blue symbols mark stars lying within the core radius.}
 \label{bailey}
 \end{center}
\end{figure}

The use of Bailey diagrams is another useful and commonly used method for defining the Oosterhoff status. In Fig.~\ref{bailey} we present Bailey diagrams for M14 in both $B$ and $V$. In this figure  we also present reference lines for OoI (\textit{solid line}) and OoII (\textit{long-dashed line}) clusters. These lines are reproduced in \citet{mz2}, and are originally from \citet{cac2005}. For the OoI component, one finds:   

\begin{align}
A_{B}^{ab} = -3.123 - 26.331 \, \log P - 35.853 \, \log P^{2},\\
A_{V}^{ab} = -2.627 - 22.046 \, \log P - 30.876 \, \log P^{2}.
\end{align}
\noindent

\noindent According to \citet{cac2005}, OoII ab-type RR Lyrae conform to the same relations,  but shifted towards longer periods by 0.06~d. We note that the shifted relation was originally fit to the distribution of what \citet{cac2005} define as \textit{evolved} stars in M3, but, as also pointed out by the authors, they also provide good reference lines for the RRab variables in several OoII clusters. In the case of RRc stars, we use the relation 

\begin{align}
A_{V}^{c} = -3.95 + 30.17 \, P - 51.35 \, P^{2} 
\end{align}

\noindent for OoI clusters \citep[from][]{2015Arellano}, and 

\begin{align}
A_{V}^{c} = -9.75 + 57.3 \, P - 80 \, P^{2}
\end{align}

\noindent for OoII clusters \citep[from][]{2013Kunder}. In order to obtain the same relations in the $B$-band, we assume a ratio between $B$ and $V$ amplitudes given by $A_{B}/A_{V}=1.29\pm0.02$ \citep{2011Dicriscienzo}.

\begin{figure}
\centering
\resizebox{\columnwidth}{!}{\includegraphics{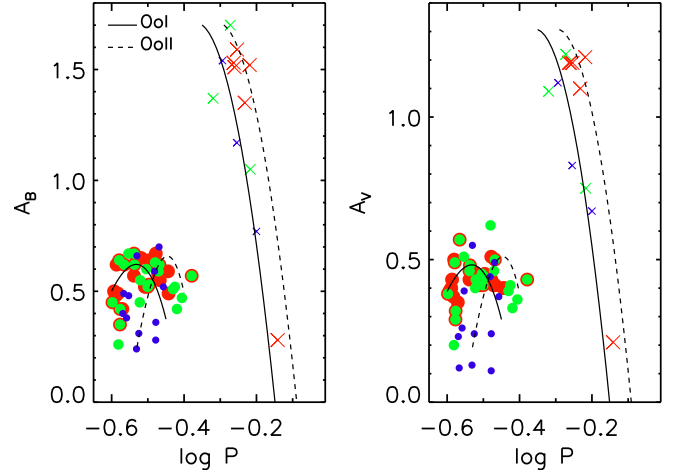}}
\caption{As in Figure~\ref{bailey}, but only for stars with $D_{m} \leq 5$. }
 \label{bailey_dmcdis}
\end{figure}

Figure~\ref{bailey} shows that RRab variables in our sample follow neither of the typical OoI or OoII lines, and that many appear to locate between the two reference loci. Some of the scatter in Bailey diagrams can arise from unidentified blends (which lead to reduced amplitudes), an effect which should become more serious as one approaches the inner, most crowded regions of a cluster. Thus, in Figure~\ref{bailey}, we mark as small symbols stars that lie inside the core radius, which is $r_c = 0.79\arcmin$ in M14. Stars with  $r> 2 \, r_c$ are shown as large symbols, and those in between are displayed with medium-sized symbols.  Although objects from the innermost part of the cluster show more scatter, it is not clear that the remainder of the sample clusters around any of the Oosterhoff lines. The possibility of an actual mix of OoI and OoII components occurring simultaneously cannot at present be excluded. In the case of RRc stars, the distribution seems, by and large, intermediate between those suggested by the corresponding OoI and OoII reference lines. 

The position of a star in the Bailey diagram can be affected by the presence of the Blazhko effect \citep{1907Blazko}; Blazhko stars show a modulation of their amplitude on a timescale that is typically much longer than the primary pulsation period. The position of a Blazhko star in a Bailey diagram can thus be misleading: unless the star is observed at the maximum-amplitude phase of the Blazhko cycle, its amplitude will appear smaller than would be the case if it were not affected by the Blazhko effect. To minimize the impact of Blazhko stars upon our analysis, we utilize the ``compatibility criterion'' discussed in Sect.~\ref{sec_fourier}, and select RRab stars with $D_{m} < 5$. These are shown as crosses in Figure~\ref{bailey_dmcdis}. Though the sample size is now greatly reduced, the distribution of RRab stars thus selected still does not clearly conform to those more typical of OoI or OoII clusters.

\begin{deluxetable*}{lccccccccc}
\tablecaption{Mean physical parameters of RR Lyrae stars in GCs.\tablenotemark{$\star$}}
\tablehead{
\colhead{Cluster} & \colhead{Oo Type} & \colhead{[Fe/H]\tablenotemark{a}} & \colhead{$M/M_{\odot}$\tablenotemark{b}} & \colhead{$\log (L/L_{\odot})$\tablenotemark{b}}  & \colhead{$T_{\rm eff}$\tablenotemark{b}}  & \colhead{$Y$\tablenotemark{b}} & \colhead{$\langle M_{V}\rangle$\tablenotemark{c,d}} & \colhead{$T_{\rm eff}^{\langle V-K \rangle}\tablenotemark{c}$} & \colhead{Reference} \\ 
 & & & & & (K) & & & (K) 
}
\startdata
NGC\,6441        & OoIII  & -0.46 &  0.47 & 1.65 & 7408 & -- & 0.68$^{{(3)}}$ & 6607 & \citet{2001Pritzl} \\
NGC\,6388        & OoIII  & -0.55 & 0.48 & 1.62 & 7495 & -- & 0.66$^{{(3)}}$ & 6607 & \citet{2002Pritzl} \\
NGC\,6362        & OoI    & -0.99 & 0.53 & 1.66 & 7429 & 0.29 & 0.66$^{{(1)}}$ & 6555 & \citet{2001Olech} \\
NGC\,6171 (M107) & OoI    & -1.02 & 0.53 & 1.65 & 7447 & 0.29 & 0.65$^{{(1)}}$ & 6619 & \citet{1997Clement} \\
NGC\,6266 (M62)  & OoI    & -1.18 & 0.53 & 1.66 & 7413 & 0.29 & 0.63$^{{(1)}}$ & 6501 & \citet{2010Contreras}\\
NGC\,6402 (M14)  & OoInt? & -1.28 &  0.53 & 1.67 & 7379 & 0.29 &0.57$^{{(2)}}$ & 6501 & This work\\
NGC\,5904 (M5)   & OoI    & -1.29 & 0.56 & 1.67 & 7362 & --& 0.58$^{{(2)}}$ & 6471 & \citet{2016Arellano}\\
NGC\,6864 (M75)  & OoInt  & -1.29 & 0.53 & 1.67 & 7399 & 0.29 &0.61$^{{(1)}}$ & 6529 & \citet{corwin_m75} \\
NGC\,6981 (M72)  & OoI    & -1.42 & -- & 1.65 & 7327& --& 0.62$^{{(2)}}$ & 6418 & \citet{2011Bramich}\\
NGC\,6229        & OoI    & -1.47 & 0.54 & 1.67 & 7362 & --& 0.62$^{{(2)}}$ & 6456 & \citet{2015Arellano}\\
NGC\,6934        & OoI    & -1.47 & 0.63 & 1.72 & 7290 & 0.27& 0.61$^{{(1)}}$ & 6455 & \citet{2001Kaluzny}\\
NGC\,5272 (M3)   & OoI    & -1.50 & 0.60 & 1.71 & 7321 & -- & 0.60$^{{(2)}}$ & 6449 & \citet{cac2005}\\
NGC\,1904 (M79)  & OoII   & -1.60 & 0.63 & 1.41 & 7293 & --& 0.41$^{{(2)}}$ & 6291 & \citet{2012Kains} \\
NGC\,7089 (M2)   & OoII   & -1.65 & 0.54 & 1.74 & 7215 & 0.27 & 0.51$^{{(1)}}$ & 6276 & \citet{2006Lazaro} \\
NGC\,5286        & OoII   & -1.69 & 0.60 & 1.72 & 7276 & 0.27& 0.52$^{{(1)}}$ & 6266 & \citet{mz2}\\
NGC\,7492        &  ?     & -1.72 & -- & -- & --& --& 0.38$^{{(2)}}$ & -- & \citet{2013Figuera} \\
NGC\,6333 (M9)   & OoII   & -1.77 & 0.5 & 1.64 & 7277 & --& 0.47$^{{(2)}}$ & 6353 & \citet{2013Arellano} \\
NGC\,4147        & OoI    & -1.80 & 0.55 & 1.69 & 7335 & 0.28& 0.60$^{{(1)}}$ & 6633 & \citet{2004Arellano}\\
NGC\,2298        & OoII   & -1.92 & 0.59 & 1.75 & 7200 & 0.26 & -- & -- & \citet{1995Clement}\\
NGC\,6809 (M55)  & OoII   & -1.94 & 0.53 & 1.75 & 7193 & 0.27& 0.53$^{{(1)}}$ & 6333 & \citet{1999Olech}\\
NGC\,5466        & OoII   & -1.98 & -- & 1.70 & 7191& --& 0.52$^{{(2)}}$ & 6328 & \citet{2008Arellano} \\
NGC\,5024        & OoII   & -2.10 & -- & 1.70 & 6282& --& 0.46$^{{(2)}}$ & 6282 & \citet{2011Arellano} \\
NGC\,4590 (M68)  & OoII   & -2.23 & 0.70 & 1.66 & 7123 & -- & -- & -- & \citet{2015Kains}\\
NGC\,5053        & OoII   & -2.27 & 0.71 & 1.69 & 7194& --& 0.49$^{{(2)}}$ & 6209 & \citet{2010Arellano} \\
NGC\,6341 (M92)  & OoII   & -2.31 & 0.64 & 1.77 & 7186 & 0.26 & 0.47$^{{(1)}}$ & 6160 & \citet{2006Lazaro} \\ 
NGC\,7099 (M30)  & OoII   & -2.34 & -- & 1.65 & 7180& --& 0.40$^{{(2)}}$ & 6227 & \citet{2013Kains} \\
NGC\,7078 (M15)  & OoII   & -2.37 & 0.76 & 1.81 & 7112 & 0.24& 0.47$^{{(1)}}$ & 6237 & \citet{2006Arellano} 
\enddata
\tablenotetext{$\star$}{As explained in the text, mass, luminosity, $Y$ and $T_{\rm eff}$ values derived from Fourier parameters should be treated with great caution, which is the reason they are not provided for each individual M14 variable in our paper. However, mean values are still listed in this table, though solely with the purpose of comparing M14 with other GCs for which such values had been previously obtained in the literature, using the same technique.}
\tablenotetext{a}{Metallicities are listed in the \citet{cg2} (UVES) scale.} 
\tablenotetext{b}{From Fourier decomposition of RRc light curves.} 
\tablenotetext{c}{From Fourier decomposition of RRab light curves.}
\tablenotetext{d}{Following \citet[][see their Table~10]{2012Kains}, the  listed absolute magnitudes, $\langle M_{V}\rangle$, are converted to values consistent with the distance scale of \citet{cac2005}. If the \citet{kovacs1998} scale is used in the original publication, then 0.2~mag are subtracted from $\langle M_{V}\rangle$. In the table, a superscript (1) marks GCs  where original values were calculated using the \citet{kovacs1998} scale, whilst (2) marks GCs where $\langle M_{V}\rangle$ was derived using the scale of \citet{cac2005}. Superscript (3) indicates cases where $\langle M_{V}\rangle$ was left the same as the original value, given that the shift suggested by \citet{cac2005} might not be applicable to metal-rich GCs \citep{2012Kains}.}
\label{tab_corwin2}
\end{deluxetable*}

All of the above points to M14 as being an OoInt cluster, a classification that much more often applies to GCs in nearby extragalactic systems than to bona-fide Milky Way halo objects \citep[see][and references therein]{2009Catelan}. In Table~\ref{tab_corwin2}, the physical parameters of M14 derived from Fourier analysis are compared for a large number of GCs of different Oosterhoff types. As can be seen from this table, the physical parameters of the M14 RR Lyrae turn out to be very similar to those of M75, which is another OoInt GC. The table suggests that the physical parameters of both M14's and M75's RR Lyrae stars are different from those of OoII or OoIII GCs. However, they do not differ much from those of OoI GCs of similar metallicity.

\begin{figure}
\begin{center}
 \resizebox{\columnwidth}{!}{\includegraphics{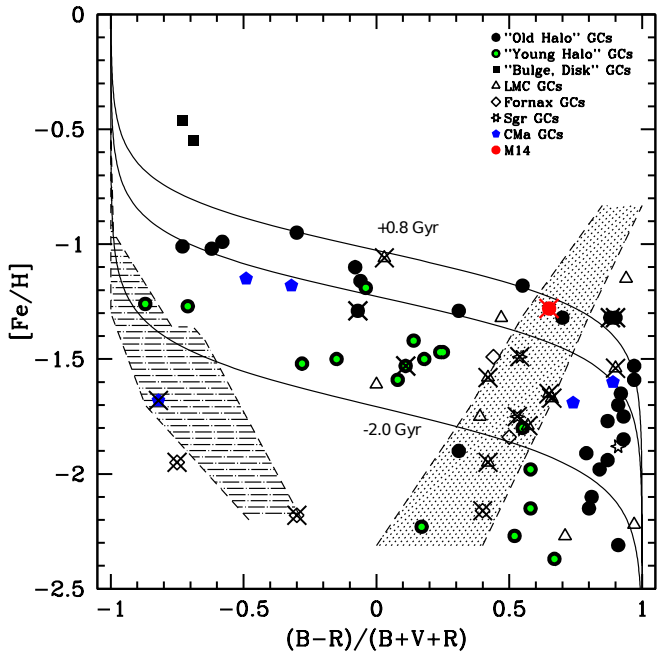}}
 \caption{${\rm [Fe/H]}-\mathcal{L}$ plane for Galactic and dSph GCs \citep[adapted from][]{2009Catelan}. Shaded areas correspond to theoretically predicted OoInt areas, from  \citet{1990Lee} (dash-dotted region) and \citet{1994Bono} (dotted region). An overplotted $\times$ symbol marks GCs for which an OoInt classification has been assigned. The solid lines represent isochrones taken from \citet{1993Catelan_dfp}.}
 \label{fehl2}
 \end{center}
\end{figure}

In \citet{2013Contreras} we studied the HB morphology of M14 through the Lee-Zinn parameter $\mathcal{L} \equiv \mathcal{(B-R)/(B+V+R)}$, where $\mathcal{B, R, V}$ are the numbers of blue, red, and variable (RR Lyrae-type) HB stars, respectively \citep{1986Zinn,1990Lee}. In Figure~\ref{fehl2} we plot the location of M14 in the ${\rm [Fe/H]}-\mathcal{L}$ plane. Even though M14 does not fall in a region of this diagram that is typically occupied by extragalactic GCs, it is interesting to note that it is located in the OoInt area that was predicted by the theoretical models of \citet{1994Bono}. The theoretical predictions of \citet{1990Lee} for redder HB types are also shown in Figure~\ref{fehl2}, where its relative success is apparent, as two Fornax GCs, along with Rup~106, all of which are OoInt clusters, fall in or close to this predicted area. We note that \citet{1990Lee} also predicts an OoInt area for bluer HB types (not shown), although this is limited to a  narrower range in $\mathcal{L}$ than is the case in \citet{1994Bono}. 

From \citet{2013Contreras}, and as also seen in \citet{piotto2002}, M14 has an extended horizontal branch (EHB).\footnote{Following the EHB classification scheme of \citet{2007Lee}.} \citet{2007Lee} found that EHB-bearing GCs show clear differences in mass and kinematics, compared to GCs without EHBs. They speculate that this difference may point to an extragalactic origin of GCs with EHBs; in particular, they might be former nuclei of dwarf galaxies that were disrupted as they were accreted by the Milky Way, or genuine GCs formed in the outskirts of the ``building blocks'' that were later accreted to the Galaxy's halo. In this sense, \citet{2007Gao} also studied the possible origin of halo GCs from past accretion events. Searching for GCs sharing both specific energy and angular momentum and that lie on the same orbital plane, as expected if they have a common origin, they found five possible ghost streams that may be associated to merging events, with M14 being a member of one of these streams. Interestingly, NGC\,6864 (M75), an EHB cluster with a possible OoInt classification \citep{corwin_m75,2006Scott}, is also a suggested member of the same stream. The remaining cluster in this ghost stream, NGC\,6535, has only two reported variables \citep{1977Liller}, one of them corresponding to an RRab star with a period of $P = 0.5884$~d (i.e., close to the middle of the ``Oosterhoff gap'' in Fig.~\ref{osdich_all}), and the other being a short-period RRc. Judging from their magnitudes, however, both of these stars are likely field stars in the background of the cluster \citep{MHL80,AS94}. Interestingly, several of the GCs belonging to other ghost streams in \citet{2007Gao} had previously been suggested to have an extragalactic origin.  

\begin{figure}
\begin{center}
 \resizebox{\columnwidth}{!}{\includegraphics{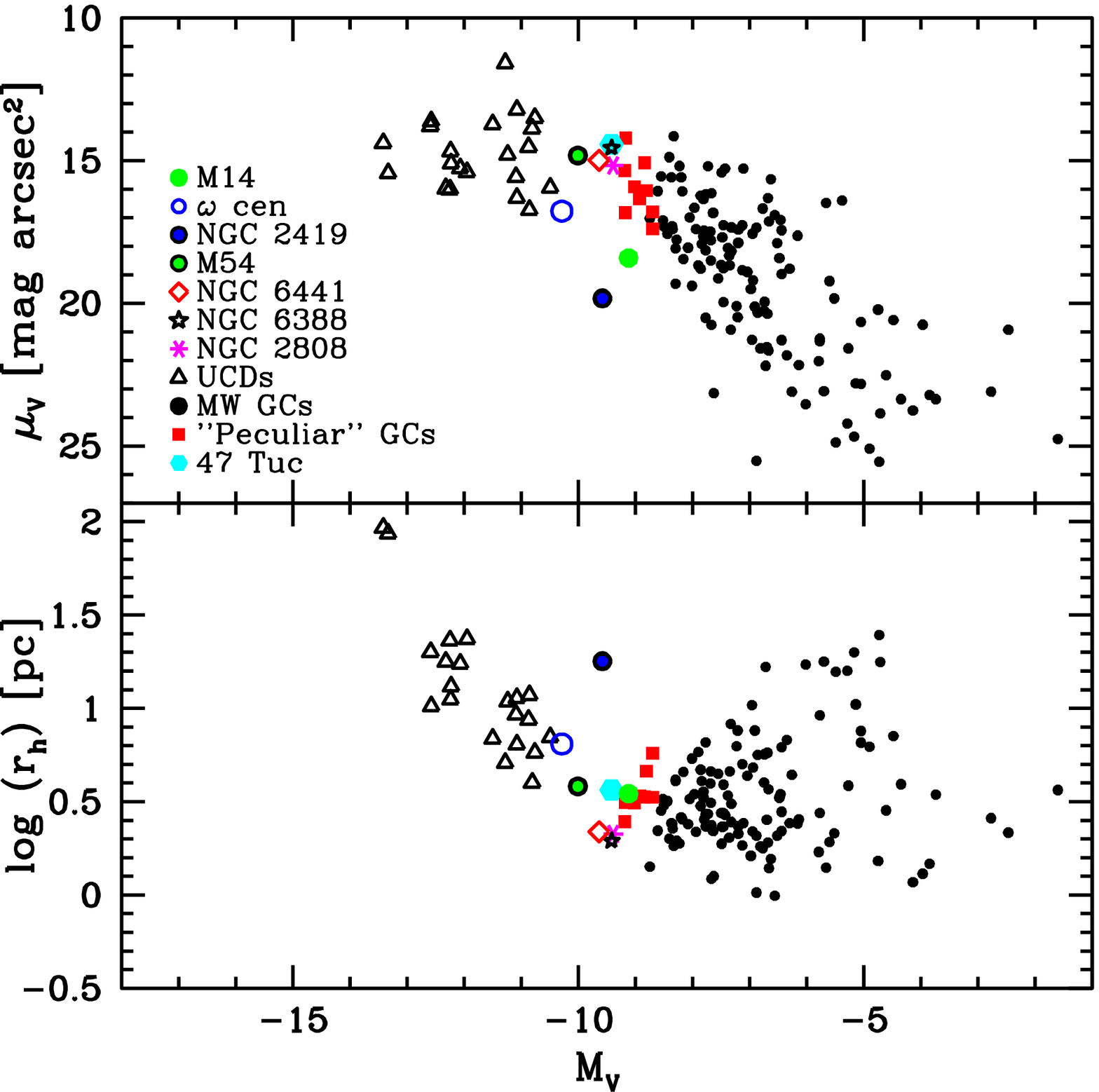}}
 \caption{Surface brightness, $\mu_{V}$, vs absolute magnitude, $M_{V}$ (\textit{top}) and $\log$ of half-mass radius, $\log r_{h}$ vs absolute magnitude, $M_{V}$ (\textit{bottom}) for Ultra Compact Dwarfs (UCDs) in Fornax and Virgo Clusters \citep[from][]{2008Evstigneeva} and GCs in the Milky Way \citep[from][]{2005Mackey,h96}.}
 \label{ucds}
 \end{center}
\end{figure}

\begin{figure}
\begin{center}
 \resizebox{\columnwidth}{!}{\includegraphics{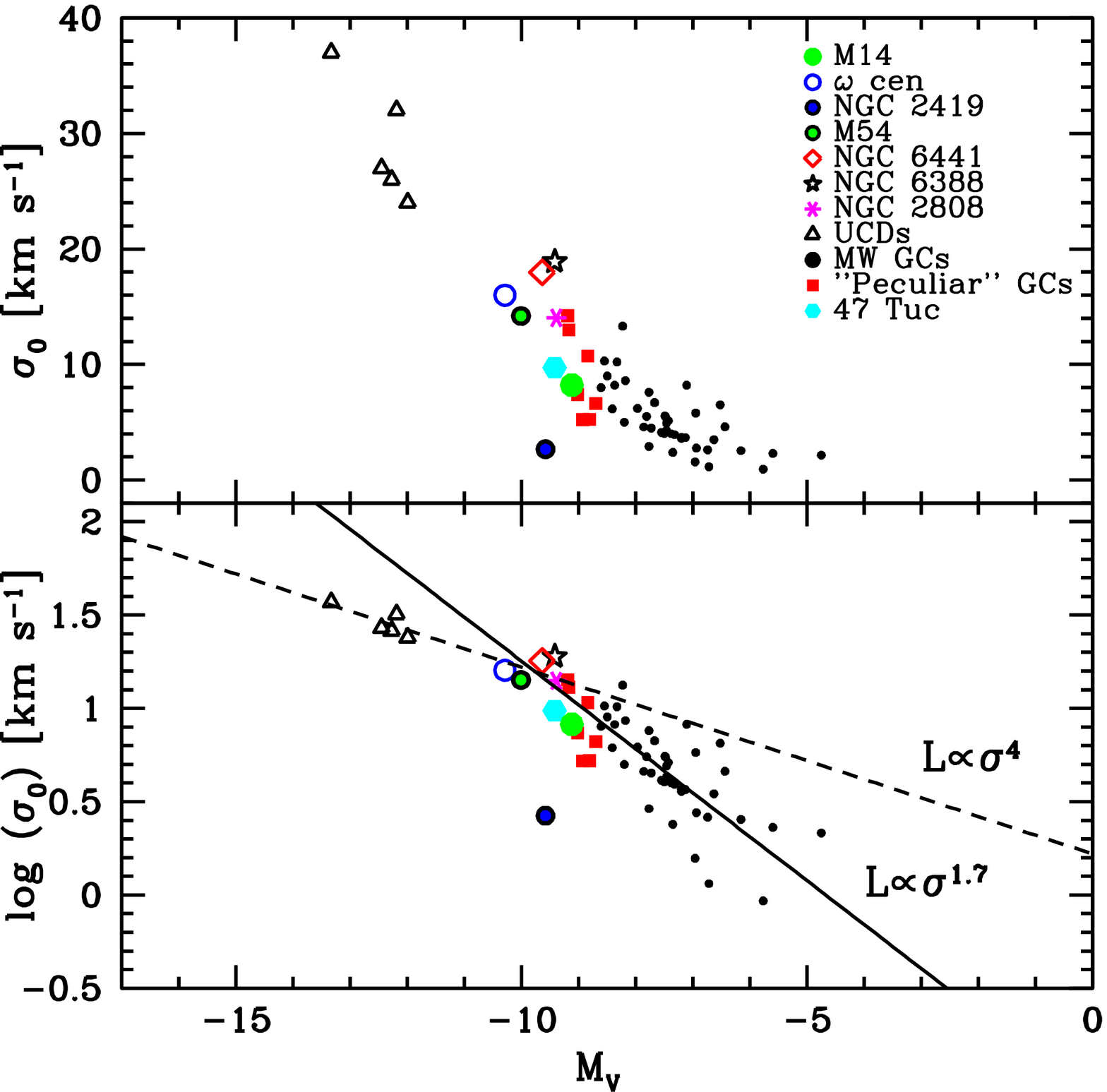}}
 \caption{Central velocity dispersion, $\sigma_{0}$, vs absolute magnitude, $M_{V}$  (\textit{top}) and $\log$ of central velocity dispersion, $\log \sigma_{0}$, vs absolute magnitude, $M_{V}$ (\textit{bottom}) for UCDs in the Fornax Cluster and GCs in the Milky Way. In the bottom panel, the Faber-Jackson relations for UCDs (dashed line) and GCs (solid line) are also overplotted.}
 \label{UCDs_vd}
 \end{center}
\end{figure}

We also note, in this context, that one of the members of M14 is a CH star \citep{cote1997}. Such stars are usually found in dSphs (see Sect.~\ref{nv24_sec}). M14 shares this unusual characteristic with $\omega$~Cen, which has been suggested to be the remnant nucleus of a dSph galaxy \citep[e.g.,][]{MAEA05,2010Carretta,JFTEA15}. Interestingly, the positions of $\omega$~Cen and M14 are also intermediate between those of GCs and ultra-compact dwarf galaxies (UCDs), in plots of integrated absolute magnitude as a function of half-mass radius $r_h$ and (especially) central surface brightness $\mu_V$ (Fig.~\ref{ucds}). In Figure~\ref{ucds} it is possible to observe that M14 is close to the bright limit of the GC luminosity distribution, alongside such bright objects as 47~Tuc, NGC\,2808, NGC\,6388, and NGC\,6441. A number of GCs are tentatively marked as ``peculiar'' in these plots, as they are not only very luminous as well, but also seem to not perfectly follow the same trends as defined by the fainter GCs.  Among these ``peculiar'' GCs one finds clusters possessing EHBs alongside several objects that have been specifically suggested to have an extragalactic origin, such as M3, M53, and NGC\,5824 \citep[e.g.,][]{2003Bellazzini,2009Newberg,2010Forbes,2010Law}.

In Figure~\ref{UCDs_vd} we plot the central velocity dispersion ($\sigma_{0}$) against absolute magnitude for 5 UCDs from the Fornax Cluster \citep[from][]{2004Drinkwater} and 57 GCs \citep[][]{1993Pryor,2008Pasquato,cote1997}; also plotted are the (extrapolated) Faber-Jackson relation of elliptical galaxies, which UCDs seem to follow \citep{2004Drinkwater}, along with the steeper relation followed by GCs \citep{1997Djorgovski}. The symbols are the same as in Figure~\ref{ucds}. We could not find information on $\sigma_{0}$ for two of the so-called peculiar clusters, namely M53 and M19. It is interesting to see that the brightest clusters seem to follow the UCDs relation, whereas M14 appears to be in an intermediate area between the relations for UCDs and GCs, though seemingly more closely conforming to the latter.

\section{Summary}\label{sec:summary}

We have obtained time-series $BVI$ CCD photometry for the GC M14. We searched for variable stars utilizing the image subtraction technique, using the ISIS v2.2 package, supported by PSF photometry using DAOPHOT\,II/ALLFRAME. Among the 165 variable stars listed in the \citet{2001Clement} catalog for M14, we were able to identify 122 stars, confirming the non-variable nature of at least 26 stars in the list. We report the discovery of 8 new variable stars, with 3 RR Lyrae, 3 LPVs, 1 DEB and 1 type II Cepheid. Several candidate RRe stars are found in the cluster, which however we argue are more likely to be short-period RRc stars.  

We have obtained physical parameters of RR Lyrae stars via Fourier analysis of their light curves, including its metallicity and distance modulus. A weighted mean of the [Fe/H] values obtained using the RRab and RRc stars gives for the cluster an ${\rm [Fe/H]}_{\rm UVES}=-1.25\pm0.18$~dex, which is in excellent agreement with the values from the CMD analysis of \citet{2013Contreras}. The RR Lyrae-based distance modulus falls in the range between $(m-M)_0=14.77$ and $15.05$~mag, depending on the technique used and zero point adopted in the calibration of RR Lyrae absolute magnitudes.  

According to the mean period of its ab-type RR Lyrae stars and the distribution of stars in the Bailey diagram, we find M14 to be an OoInt GC. While the number fraction of RRc stars may be closer to the value expected for an OoII cluster, we also note that this is a poor indicator of the Oosterhoff type. The physical parameters of the M14 RR Lyrae stars, derived from Fourier fitting of the light curves, are similar to those of M75, another OoInt GC with a metallicity similar to M14's. 

Since OoInt GCs are more commonly found in nearby satellites of the Milky Way than in the Milky Way proper, we discuss some of the characteristics of M14 that could point to an extragalactic origin of the cluster. We find that M14 has some properties that are intermediate between UCDs and the bulk of Milky Way GCs. UCDs have been suggested to be the nuclei of former dSphs, and several GCs that have been suggested to be of an extragalactic origin are also found to have some properties intermediate between UCDs and GCs. We review additional evidence that M14 may have originated outside the Galaxy, and conclude that the possibility cannot be discarded at present that M14 may be the massive remnant of the merger event between a dwarf galaxy and the Milky Way.

\acknowledgments
We warmly thank an anonymous referee, whose comments have greatly helped improve our paper.
Support for this project is provided by the Ministry for the Economy, Development, and Tourism's Millennium Science Initiative through grant IC\,120009, awarded to the Millennium Institute of Astrophysics (MAS); by Proyecto Basal PFB-06/2007; by FONDECYT grant \#1171273; and by CONICYT's PCI program through grant DPI20140066.

\bibliography{ref-MC3}

\appendix 

\section{Notes on individual stars}

Table~\ref{varnotes} provides information on some of the variable stars detected in previous studies (\citealt{sharam14}; WF94; CDLM12). 

In what follows, we also provide additional information on some noteworthy variables that were detected in the course of our analysis. 

\begin{deluxetable*}{lll}
\tablecaption{Notes on individual variable stars}
\tablehead{\colhead{ID} & \colhead{Study} & \colhead{Note} }
\startdata 
V26, V25, V40, & WF94 &  These 22 variable stars have notes stating that they are either blended, present small variations\\
V52-V54, V63-V67, & & and/or large scatter in WF94. These are not detected by CDLM12\\
V69, V72, V81-V87, & & and are also not found to be variable in our study. \\
V89, V93 & & \\ 
V25 & WF94 & CDLM12 cannot confirm its variability; it is detected in our study.  \\
 & & The star is at the edge of the field of view of CDLM12. \\
V27  & WF94 & Outside of our field-of-view.\\
V28  & WF94 & Outside of our field-of-view.\\
V45  & WF94 & CDLM12 cannot confirm its variability; it is detected in our study.\\
V50  & WF94 & Marked as non-variable in \citet{2001Clement}. WF94 state that the star is too blended to \\ 
& &  measure. Our V169 is 2.2\arcsec~north-east from the position of V50 in the finding chart of WF94. \\ 
& &  However, we cannot confirm that V169 is in fact the same star as V50.\\
V68  & WF94 & In our study. CDLM12 cannot confirm its variability; however, \citet{2001Clement} note that this is their V134.\\
V71  & WF94 & In our study. CDLM12 cannot confirm its variability; however, \citet{2001Clement} note that this is their V35.\\
V77  & WF94 &  WF94 classify this star as an RRc pulsator with a period of 0.327~d. We detect a variable star at \\
& &  the coordinates of V77, but our light curve shows that it is an ab-type RR Lyrae with a period of 0.7932~d.\\
V86 & WF94 &  \citet{2001Clement} note that this is V151 in CDLM12. We also detect this object at the coordinates of V151.\\
& &  However, the position does not agree with the location of V86 in Fig.~1b of WF94, which the authors \\
& & claim as the correct position for this star. Thus we confirm that V86 is not a variable star.\\
V91 & WF94 & Marked as non-variable in \citet{2001Clement} and CDLM12. However, we detect variability\\
& &  at the position of V91 in Fig.~1b of WF94. This object is marked as V100 in CDLM12.\\ 
V92 & WF94 & Marked as non-variable in \citet{2001Clement} and CDLM12. However, we detect variability\\
& &  at the position of V92 in Fig.~1b of WF94. This object is marked as V109 in CDLM12.\\
Nova & \citet{sharam14} & Nova Ophiuchi 1938. Not detected in our study. \\ 
V111 & CDLM12 & The coordinates of this object are close to our V172. However, V111 has a different period and classification\\
& & than V172. The low-amplitude variability of V111 probably makes it hard to detect in our study.\\
V94 & CDLM12 & Low-amplitude, classified as an RRc variable star. It is not detected in our study. \\
V108 & CDLM12 & Low-amplitude, classified as an RRc variable star. It is not detected in our study. \\
V113 & CDLM12 & Low-amplitude, classified as an RRc variable star. It is not detected in our study. \\
V115 & CDLM12 & Low-amplitude, classified as an RRc variable star. It is not detected in our study. \\
V101 & CDLM12 & Long-period variable star. It is not detected in our study.\\
V103 & CDLM12 & Long-period variable star. It is not detected in our study.\\
V125 &  CDLM12 & Long-period variable star. It is not detected in our study.\\
V146 & CDLM12 & Long-period variable star. It is not detected in our study.\\
V155 & CDLM12 & Long-period variable star. It is not detected in our study.\\
V156 & CDLM12 & Long-period variable star. It is not detected in our study.\\
V161 & CDLM12 & Listed as an SX~Phe variable star, this object is not detected in our study. 
\enddata
\label{varnotes}
\end{deluxetable*} 

\subsection{V97} 

\begin{figure}[hb]
\begin{center}
 \plotone{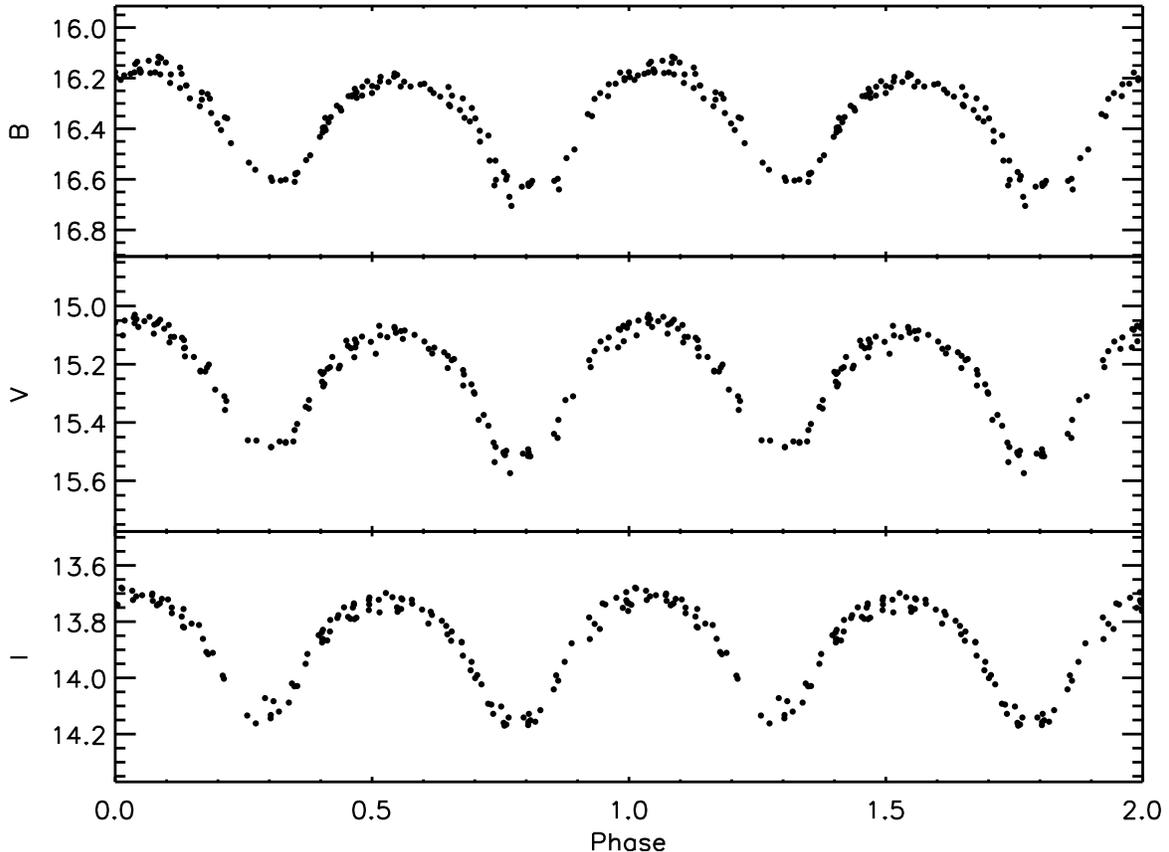}
 \caption{$B$, $V$, $I$ (from top to bottom) light curves for the W~UMa star V97.}
 \label{nv52_bvi}
 \end{center}
\end{figure}

Figure~\ref{nv52_bvi} shows the $B$, $V$, $I$ light curves of V97. Here we observe a sinusoidal-like variation, with a period of $P=0.37744$~d, with the presence of two minima that are well defined but do not differ greatly in depth. The short period and light curve morphology alike suggest that V97 is a W~UMa-type (contact) eclipsing binary.

W~UMa stars are expected to fall 4-5 magnitudes below the RR Lyrae level \citep{2000Ruc}. However, inspection of the CMD (Fig.~\ref{cmd1}) shows that V97 is instead found 2~mag above the HB level in $V$. This strongly suggests that the star is not a cluster member.  

To confirm this, we use the membership method of \citet{2000Ruc}, where the absolute magnitude derived from $M_{V}^{\rm obs} = V-(m-M)_{V}$ is compared against the star's expected absolute magnitude, given its period and color. The latter is given by \citep{1994Ruc}

\begin{equation}
 M_{V}^{cal} = -4.44 \, \log P + 3.02 \, (\bv)_{0} + 0.12. 
\end{equation}

\noindent \citet{1994Ruc} classifies such a variable star as a cluster member if $| \Delta M_{V} |<1.0$. Deviations larger than 1~mag imply that the binary is most likely a field star.

In the case of V97, we use our favored value of $(m-M)_{V} = 16.63$~mag (see Sect.~\ref{sec:DM}) and $E(B-V) = 0.57$. From our observations, $\langle V \rangle = 15.309$~mag and $(\bv)=1.134$~mag, which yields a difference of $| \Delta M_{V} | = 5.03$~mag. Thus, it is safe to conclude that V97 is not a member of M14, but instead a field W~UMa star.

\subsection{V165}\label{nv24_sec}

V165 is one of the newly discovered variables and it is classified as an LPV based on its light curve and its position in the CMD, falling at the tip of the RGB. \citet{cote1997} reports the discovery of a CH star in M14. Comparison of their finding chart with the position of our V165 allows us to identify this new variable as the reported CH star. 

Classical CH stars are binary stars comprised of a red-giant primary and a white-dwarf secondary, with the peculiar abundances of the former resulting from mass transfer via stellar winds or Roche lobe overflow during the ascent of the white dwarf progenitor up the AGB. It is thought that environment plays a key role in the evolution of this type of stars, making CH stars more likely to exist in low-concentration systems such as M14. Consistent with this notion, \citet{1977McClure} pointed out that $\omega$~Cen and M22, two GCs possessing CH stars, are relatively less concentrated and fall away from clusters of similar mass in the $\log c - M_{V}$ plane, with $c$, the cluster concentration, defined as $c = r_{t}/r_{c}$.

CH stars have also been reported in several dSph galaxies such as Ursa Minor, Draco, Sculptor, Carina, in addition to $\omega$~Cen \citep[e.g.][]{1984McClure,1996Mayor,cote1997}. NGC\,2419, another massive and relatively low-concentration system, has been suggested to be profitable to search for these kinds of stars \citep{1977McClure}. It is noteworthy that both $\omega$~Cen and NGC\,2419 have also been suggested to be the cores of disrupted dSph galaxies \citep[e.g.][]{2003Bekki,2004Vdb_mackey}. As discussed in Sect.~\ref{sec:oost}), this provides additional evidence that M14 too may have originated outside of the Milky Way proper.

\subsection{V167 and the Cepheid Distance Modulus}

We report the discovery of a new type II Cepheid, V167, with a period of 6.20~d and a visual amplitude of  $A_{V} = 0.26$~mag. 

Type II Cepheids have periods in a more limited range (between about 1 and 80~d) than the classical Cepheids, and have lower masses, metallicities, and luminosities than their classical counterparts \citep[e.g.,][]{CS15}. They also differ in their period-luminosity relation, with type II Cepheids being, on average, about 1.5~mag fainter than classical Cepheids with the same period. 

While the period  of V167 falls in the typical range of W~Vir stars, its amplitude falls below the typical values for these variables (0.3-1.5~mag). However, we note that it is not clear that we have covered the minimum light. The PDM periodogram also shows a large uncertainty in the period, due to our incomplete phase coverage.

\begin{deluxetable}{ccccc}
\tablecaption{Type II Cepheids in M14}
\tablehead{
\colhead{ID} & \colhead{$P$ (d)} & \colhead{$\langle V \rangle$} & \colhead{$M_{V}$} & $(m-M)_{V}$
}
\startdata
V1 & 18.729 & 14.227 & -2.04 $\pm$ 0.08 & 16.27 $\pm$ 0.08\\
V2 & 2.79471 & 15.704 & -0.68 $\pm$ 0.05 & 16.39 $\pm$ 0.05\\
V7 & 13.6038 & 14.748 & -1.81 $\pm$ 0.08 & 16.56 $\pm$ 0.08\\
V17 & 12.091 & 14.750 &  -1.73 $\pm$ 0.07 & 16.48 $\pm$ 0.07\\
V76 & 1.89026 & 16.017 & -0.40 $\pm$ 0.05 & 16.42 $\pm$ 0.05\\
V167 & 6.201018 & 15.576 & -1.25  $\pm$ 0.06 & 16.83 $\pm$ 0.06\\
\hline
Mean & &  & & 16.49 $\pm$ 0.17
\enddata
\label{tdist_cep}
\end{deluxetable}

In addition to V167, M14 possesses 5 more type II Cepheids, some of whose properties are summarized in Table~\ref{tdist_cep}. Using the magnitude-weighted mean magnitude of type II Cepheids in M14 and the period-luminosity relation of \citet{2003Pritzl}, we are in a position to obtain a distance modulus of the cluster that is independent of the RR Lyrae stars discussed in Sect.~\ref{sec:DM}. In Table~\ref{tdist_cep} we can observe a large scatter between the individual values of the distance modulus. The latter could be explained by the fact that \citet{2003Pritzl} derive the period-luminosity relation for type II Cepheids in GCs NGC~6441 and NGC~6388, which are known to be peculiar objects \citep[][and references therein]{CS15}. Although the authors consider this relation valid for other Milky Way GCs, Fig.~9 in \citet{2003Pritzl} does show a considerable scatter among type II Cepheids in Galactic GCs.

From Table~\ref{tdist_cep} we find $(m-M)_{V}=16.49 \pm 0.17$~mag (standard deviation of the mean), which is consistent with the value from RRab stars and with the listed value in the \citet{h96} catalog. While this is 0.2~mag shorter than the value favored in Sect.~\ref{sec:DM}, the two values are formally consistent with one another, to within their respective errors.

\subsection{V168}

\begin{figure}
\centering
\epsscale{0.75}
\plotone{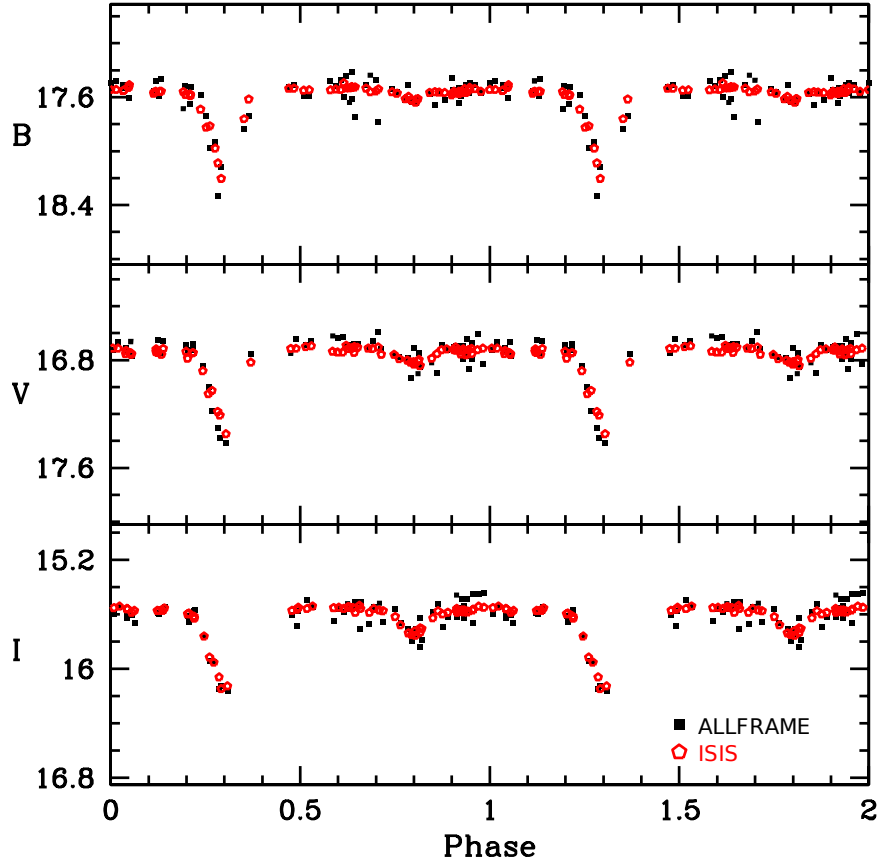}
\caption{$B$, $V$, $I$ (from top to bottom) light curves for V168, a DEB. Black and red symbols correspond to the ALLFRAME and ISIS reductions, respectively. }
\label{nv55_bvi}
\end{figure}

We report the discovery of a DEB (Algol-type eclipser), designated as V168, with a period of $P \sim 1.27$~d. The DAOPHOT\,II/ALLFRAME and ISIS light curves are presented in Figure~\ref{nv55_bvi}, where it is possible to observe that the ISIS light curve presents significantly less scatter than the ALLFRAME one, as frequently happens in such crowded fields. We derive the period with PDM using the ISIS data, finding $P \simeq 1.27$~d.

The location of the star on the CMD  (Fig.~\ref{cmd1}) is rather interesting as in principle the binary could result from the combination of an HB star with an RGB star. However, the short period of the binary rules out this possibility, in view of the large diameters of HB and (especially) RGB stars. We point out that more observations are needed to constrain the period, as the primary eclipse in our observations is only covered partially (Fig.\ref{nv55_bvi}). 

\end{document}